\documentclass[%
 reprint,
 amsmath,amssymb,
 aps,
 nofootinbib
]{revtex4-1}

\usepackage{amsthm}
\usepackage[dvipdfmx]{graphicx}
\usepackage[dvipdfmx]{xcolor}
\usepackage{dcolumn}
\usepackage{bm}
\usepackage{braket}
\usepackage{amsmath}
\usepackage{amssymb}
\usepackage{amsthm}
\usepackage{enumerate}
\usepackage{booktabs}
\usepackage{txfonts}
\usepackage[update,prepend]{epstopdf}
\usepackage[english]{babel}
\usepackage{listings}
\usepackage{qcircuit}
\usepackage[subrefformat=parens]{subcaption}
\usepackage{caption}
\usepackage{multirow}
\usepackage{algorithmic}
\usepackage{algorithm}
\usepackage{here}
\newtheorem{assumption}{Assumption}
\newtheorem{theorem}{Theorem}

\newtheorem{corollary}{Corollary}

\allowdisplaybreaks[1]

\makeatletter
\def\hlinewd#1{%
	\noalign{\ifnum0=`}\fi\hrule \@height #1 %
	\futurelet\reserved@a\@xhline}
\makeatother

\begin{document}

\preprint{APS/123-QED}

\title{Bermudan option pricing by quantum amplitude estimation and Chebyshev interpolation}

\author{Koichi Miyamoto}
\email{koichi.miyamoto@qiqb.osaka-u.ac.jp}
\affiliation{Center for Quantum Information and Quantum Biology, Osaka University \\ 1-3 Machikaneyama, Toyonaka, Osaka, 560-8531, Japan}

\date{\today}

\begin{abstract}

Pricing of financial derivatives, in particular early exercisable options such as Bermudan options, is an important but heavy numerical task in financial institutions, and its speed-up will provide a large business impact.
Recently, applications of quantum computing to financial problems have been started to be investigated.
In this paper, we first propose a quantum algorithm for Bermudan option pricing.
This method performs the approximation of the continuation value, which is a crucial part of Bermudan option pricing, by Chebyshev interpolation, using the values at interpolation nodes estimated by quantum amplitude estimation.
In this method, the number of calls to the oracle to generate underlying asset price paths scales as $\widetilde{O}(\epsilon^{-1})$, where $\epsilon$ is the error tolerance of the option price.
This means the quadratic speed-up compared with classical Monte Carlo-based methods such as least-squares Monte Carlo, in which the oracle call number is $\widetilde{O}(\epsilon^{-2})$.

\end{abstract}

\pacs{Valid PACS appear here}
                              
\maketitle

\section{\label{sec:intro}Introduction}

Following the recent advances of quantum computing technologies\footnote{As a standard textbook for quantum computing, we refer to \cite{NC}.}, many researches have been done for the application to practical problems in various industries.
One of the promising targets is finance (see \cite{Orus,Egger,Bouland} for reviews).
Financial firms have a lot of heavy computational tasks in their daily business\footnote{As standard textbooks for financial engineering, especially option pricing, we refer to \cite{Hull,Shreve}.}, and therefore the speed-up of such tasks by quantum computers are expected to provide a large impact.
For example, previous papers studied option pricing~\cite{Rebentrost,Martin,Stamatopoulos,Ramos-Calderer,Fontanela,Vazquez,Kaneko,Tang,Chakrabarti,An,Gonzalez-Conde,Radha}, risk measurement~\cite{Worner,Egger2,Miyamoto2,Kaneko1}, portfolio optimization~\cite{Rebentrost2,Kerenidis1,Hodson}, and so on.

In this paper, we focus on {\it Bermudan option pricing} and consider how to speed it up by quantum algorithms.
Let us briefly describe the problem.
An {\it option} is a financial contract between two parties, the option buyer and seller, which conveys the option buyer the right to buy some {\it underlying assets} such as stocks and bonds from or sell them to the option seller, at some specified price on some date.
Or, more generally, it can be regarded as a contract, in which the option buyer receives some amount of money ({\it payoff}) determined in reference to the underlying asset price, from the option seller.
There are some kinds of options with respect to timing of exercise of right.
In an {\it European} option, the option buyer can exercise the right at one predetermined date, which is called the {\it maturity}.
On the other hand, there are {\it early-exercisable} options, in which the option buyer can choose the exercise date.
In an {\it American} option, the option buyer can exercise the right at any time before the final maturity $T$.
In a {\it Bermudan} option, exercise of right is possible on any of finite predetermined dates including $T$.
We hereafter call such dates {\it exercise dates}.

Major financial firms hold large portfolios of a wide variety of options, and therefore pricing them is an important task for business.
However, it is also a difficult task.
Basically, the option price is the expected value of the payoff under some stochastic model describing random time evolution of the underlying assets.
Although European options can be sometimes priced easily, for example, by some analytic formulas, pricing Bermudan and American options typically involves heavy numerical calculations.
The difficulty partly stems from the nature of the problem as dynamic programming.
That is, pricing early-exercisable options contains determining the {\it optimal exercise time} as a crucial part.
Although there are some kinds of methods which aim to reflect early exercise to the option price, each one has pros and cons.

One major category of pricing methods is the Monte Carlo-based method\footnote{We refer to \cite{Glasserman} as a textbook on Monte Carlo simulation and its application to finance.}, in which we generate many sample paths of evolution of underlying asset prices, and estimate the expected payoff as an averaged payoff over the paths.
This approach has an advantage in the case of {\it multiple underlying assets}.
Namely, in this approach, the estimation error on the option price decays as $\widetilde{O}(N^{-1/2})$\footnote{$\widetilde{O}(\cdot)$ hides logarithmic factors in the ordinary big O notation $O(\cdot)$.} when the sample number $N$ increases, regardless of the number of the underlying assets $d$.
In other words, it suffices to take $\widetilde{O}(\epsilon^{-2})$ samples in order to achieve the error tolerance $\epsilon$ on the option price.
This contrasts to other methods, for example approaches based on solving partial differential equations~\cite{Tavella,Duffy}, whose complexity is $\widetilde{O}((1/\epsilon)^{{\rm poly}(d)})$.
On the other hand, in the Monte Carlo-based methods, it is difficult to precisely determine the optimal exercise time, and we have to approximate this in some way.
In many cases, this is done through approximation of the {\it continuation value}, which is the option price at each exercise date in the case that the option buyer forgoes the exercise.
The option should be exercised if the payoff is larger than the continuation value, and should not be exercised otherwise.

In this category, the {\it least-squares Monte Carlo} (LSM) \cite{Longstaff} is widely used.
LSM estimates the continuation value at each exercise date by {\it linear regression} using the generated sample paths as training data, and then, going backward from the final maturity to the present, finds the present option price.

Note that this method can also price American options approximately, replacing exercisability at any point in the continuous time period with that at discrete dates with sufficiently small intervals. 

In this paper, we propose a new method for Bermudan option pricing, combining {\it Chebyshev interpolation} and quantum algorithm for Monte Carlo integration~\cite{Montanaro,Suzuki,Herbert}, which is based on quantum amplitude estimation (QAE)~\cite{Brassard,Suzuki,Aaronson,Grinko,Nakaji,Brown,Tanaka,Kerenidis2,Uno,Giurgica-Tiron,Wang}.
As far as the author knows, this is the first proposal on the quantum method for Bermudan option pricing.
Chebyshev interpolation is a widely used method for function approximation\footnote{See \cite{Trefethen} as a textbook on this topic}, and has already been used in some (classical) methods for Bermudan option pricing~\cite{Sullivan,Lim,Mahlstedt,Gas,Glau1,Glau2,Glau3}.
In the proposed method, given the access to the quantum circuit (or, the {\it oracle}) for time evolution of underlying asset prices, we calculate the continuation values at the interpolation nodes by the quantum algorithm, and find Chebyshev interpolation using these values.
Importantly, this method outputs an estimation of the option price with the error at most $\epsilon$, calling the oracle only $\widetilde{O}(\epsilon^{-1})$ times.
Thus, as we commonly observed in applications of QAE to various kinds of problems, we obtain the {\it quadratic speed-up} compared with the classical Monte Carlo-based methods such as LSM and the Chebyshev interpolation-based methods.

The rest of this paper is organized as follows.
In Section \ref{sec:Cheb}, we briefly explain Chebyshev polynomials and function approximation by them.
We present how to calculate the coefficients of Chebyshev expansion in general and the upper bound for the approximation error.
In Section \ref{sec:Berm}, we present the general formulation of Bermudan option pricing and explain LSM as a typical classical solutions.
In Section \ref{sec:QAE}, we explain QAE and QAE-based Monte Carlo integration.
Then, in Section \ref{sec:NewAlgo}, we present the new algorithm for Bermudan option pricing based on Chebyshev interpolation and QAE.
We also present an upper bound on the price error in the method, and that on the complexity sufficient to achieve the given error tolerance.
Section \ref{sec:Sum} summarizes this paper.
All proofs are presented in the appendix.

\subsection{Notations}

We here explain the notations used in this paper.

$\mathbb{N}$ denotes the set of all positive integers, and $\mathbb{N}_0 := \{0\}\cup\mathbb{N}$.
We define $[n]:=\{i\in \mathbb{N} \ | \ i\le n\}$ for any $n\in \mathbb{N}$, and $[n]_0:=\{i\in \mathbb{N}_0 \ | \ i\le n\}$ for any $n\in \mathbb{N}_0$.
We also define $\mathbb{N}_{\ge n}:=\{i\in \mathbb{N} \ | \ i\ge n\}$ for $n\in \mathbb{N}$.
Similarly, we define $\mathbb{R}_{>a}:=\{x\in\mathbb{R} \ | \ x>a\}$ and $\mathbb{R}_{\ge a}:=\{x\in\mathbb{R} \ | \ x\ge a\}$ for $a\in\mathbb{R}$.
$\mathbb{R}_+$ denotes the set of all positive real number, that is, $\mathbb{R}_{>0}$.


For $a,b\in \mathbb{N}_0$, $\delta_{a,b}$ denotes the Kronecker delta, which is 1 if $a=b$ or 0 otherwise.
For $d\in\mathbb{N}$ and $\vec{l}_1,\vec{l}_2\in \mathbb{N}_0^d$, we also define $\delta_{\vec{l}_1,\vec{l}_2}$, which is 1 if $\vec{l}_1=\vec{l}_2$ and 0 otherwise.

For a measure space $(\Omega,\mathcal{F},\mu)$ and $p\in \mathbb{R}_{\ge 1}$, $L^p(\Omega,\mu)$ denotes the $L^p$ space on it.


The indicator function $1_C$ takes 1 if the condition $C$ is satisfied, and 0 otherwise.

In this paper, we consider quantum states of systems consisting of some quantum registers with some qubits.
For $x\in\mathbb{R}$, $\ket{x}$ denotes one of the computational basis states on some register, whose bit string corresponds to the binary representation of $x$ with truncation at some digit.
For $d\in\mathbb{N}$ and $\vec{x}=(x_1,...,x_d)^T\in\mathbb{R}^d$, $\ket{\vec{x}}:=\ket{x_1}...\ket{x_d}$ is the state on the $d$-register system.

\section{\label{sec:Cheb}Approximation of functions by Chebyshev interpolation}

For $l\in \mathbb{N}_0$, the $l$-th Chebyshev polynomial (of the first kind) is defined as
\begin{equation}
	T_l(x):=\cos(l\cdot\arccos(x)),
\end{equation}
where $x\in[-1,1]$.
One of its important properties is the {\it discrete} orthogonality: for any $m\in \mathbb{N}_0$ and $l_1,l_2\in[m]_0$,
\begin{equation}
	\sum_{j=0}^{m} T_{l_1}(x_{m,j})T_{l_2}(x_{m,j})=
	\begin{cases}
		0 &; \ {\rm if} \ l_1\ne l_2 \\
		m+1 &; \ {\rm if} \ l_1=l_2=0 \\
		\frac{m+1}{2} &;  \ {\rm if} \ l_1=l_2>0 \\
	\end{cases}. \label{eq:discOrth}
\end{equation}
Here, $x_{m,j}$ is the {\it Chebyshev node} defined as
\begin{equation}
	x_{m,j} = \cos\left(\frac{j+\frac{1}{2}}{m+1}\pi\right)
\end{equation}
for $j\in[m]_0$.
$x_{m,0},...,x_{m,m}$ are the zeros of $T_{m+1}(x)$.

We also define the {\it tensorized} Chebyshev polynomials on a general hyperrectangle in $\mathbb{R}^d$, where $d\in\mathbb{N}$.
That is, given
\begin{equation}
	\mathcal{D}=[L_1,U_1]\times...\times[L_d,U_d], \label{eq:hypRect}
\end{equation}
with $L_1,...,L_d,U_1,...,U_d\in\mathbb{R}$ satisfying $L_1<U_1,...,L_d<U_d$, we define
\begin{equation}
	\widetilde{T}_{\mathcal{D},\vec{l}} \ (\vec{S}) := \prod_{i=1}^d T_{l_i}\left(\theta_{\mathcal{D},i}(S_i)\right)
\end{equation}
for every $\vec{l}=(l_1,...,l_d)^T\in \mathbb{N}_0^d$ and $\vec{S}=(S_1,...,S_d)^T\in\mathcal{D}$, where 
\begin{equation}
\theta_{\mathcal{D}}(\vec{S}) := \left(\frac{2S_1-U_1-L_1}{U_1-L_1},...,\frac{2S_d-U_d-L_d}{U_d-L_d}\right)^T.
\end{equation}
For the above polynomials, the orthogonality relation is now
\begin{equation}
	\sum_{\vec{x}\in\mathcal{G}^{d,m}_{\mathcal{D}}} \widetilde{T}_{\mathcal{D},\vec{l}_1}(\vec{s})\widetilde{T}_{\mathcal{D},\vec{l}_2}(\vec{s})=
	\begin{cases}
		\frac{(m+1)^d}{2^{\aleph\left(\vec{l}_1\right)}} &; \ {\rm if} \ \vec{l}_1= \vec{l}_2 \\
		0 &; \ {\rm if} \ \vec{l}_1\ne \vec{l}_2
	\end{cases} \label{eq:discOrthTen}
\end{equation}
for every $m\in\mathbb{N}$ and $\vec{l}_1,\vec{l}_2\in[m]_0^d$, where $\aleph\left(\vec{l}\right) := \# \{i\in[d] \ | \ l_i>0 \}$ for $\vec{l}=(l_1,...,l_d)^T\in \mathbb{N}_0^d$, and $\mathcal{G}^{d,m}_{\mathcal{D}}$ is the set of points $\vec{S}^{\mathcal{D},m}_{\vec{j}}\in\mathcal{D}$ written in the form of
\begin{eqnarray}
	&&\vec{S}^{\mathcal{D},m}_{\vec{j}} := \nonumber \\ &&\quad \left(\frac{U_1-L_1}{2}x_{m,j_1}+\frac{U_1+L_1}{2},...,\frac{U_d-L_d}{2}x_{m,j_d}+\frac{U_d+L_d}{2}\right)^T \nonumber \\
	&&\label{eq:nodeMod}
\end{eqnarray}
with $\vec{j}=(j_1,...,j_d)^T\in [m]_0^d$.

We can use the above polynomials for function approximation.
Given $\mathcal{D}$ as (\ref{eq:hypRect}) and $m\in\mathbb{N}$, we define the {\it Chebyshev interpolation} of a function $f:\mathcal{D}\rightarrow\mathbb{R}$ as
\begin{equation}
	\Pi_{\mathcal{D},m}[f](\vec{S}) := \sum_{\vec{l}\in [m]_0^d} a_{f,\vec{l}} \widetilde{T}_{\mathcal{D},\vec{l}} \ (\vec{S}) \label{eq:ChebExp}
\end{equation}
for every $\vec{S}\in\mathcal{D}$, where the coefficient $a_{f,\vec{l}}$ is calculated by
\begin{equation}
	a_{f,\vec{l}} := \frac{2^{\aleph\left(\vec{l}\right)}}{(m+1)^d}\sum_{\vec{S}\in \mathcal{G}^{d,m}_{\mathcal{D}}} f(\vec{S})\widetilde{T}_{\mathcal{D},\vec{l}} \ (\vec{S}) \label{eq:coef0}
\end{equation}
for every $\vec{l}\in[m]_0^d$.
This is in fact an interpolation, since $\Pi_{\mathcal{D},m}[f](\vec{S})=f(\vec{S})$ for every node $\vec{S}\in\mathcal{G}^{d,m}_{\mathcal{D}}$.

The error in the above approximation has been investigated in \cite{Gas,Sauter}.
They gave the error bound, making an assumption on analyticity of the interpolated function $f$.
We here present the theorem on such a error in the case where we are given the values of $f$ at the Chebyshev nodes with some errors \cite{Gas}.
However, let us make some definitions prior to the theorem.
For $\rho\in \mathbb{R}_{>1}$, the Bernstein ellipse $\mathcal{B}_{\rho}$ is defined as the open region in the complex plane bounded by the ellipse $\left\{\frac{1}{2}\left(u+u^{-1}\right) \ \middle| \ u\in\mathbb{C}, |u|=\rho \right\}$.
We also define the generalized Bernstein ellipse as $\mathcal{B}_{\mathcal{D},\rho} := \left(\eta_1\circ\mathcal{B}_{\rho}\right)\times\cdots\times\left(\eta_d\circ\mathcal{B}_{\rho}\right)$, where, for every $i\in[d]$, $\eta_i(z)$ is the map from $\mathbb{C}$ to $\mathbb{C}$ defined as $\eta_i(z):=\frac{U_i-L_i}{2}z+\frac{U_i+L_i}{2}$.
Furthermore, we define the multivariate version of the {\it Lebesgue constant} of the Chebyshev nodes: for every $m\in\mathbb{N}$,
\begin{equation}
	\Lambda_{d,m}:=\max_{(x_1,...,x_d)^T\in[-1,1]^d} \sum_{(j_1,...,j_d)^T\in[m]_0^d}\prod_{i=1}^d\ell^m_{j_i}(x_i),
\end{equation}
where
\begin{equation}
\ell^m_j(x):=\prod_{k\in[m]_0\setminus \{j\}} \frac{x-x_{m,k}}{x_{m,j}-x_{m,k}}
\end{equation}
for every $j\in[m]_0$.
As \cite{Gas} showed,
\begin{equation}
	\Lambda_{d,m} \le \prod_{i=1}^d \left(\frac{2}{\pi}\log(m+1)+1\right)
\end{equation}
holds, which is derived from the well-known upper bound $\Lambda_{1,m} \le \frac{2}{\pi}\log(m+1)+1$ \cite{Trefethen}.
Then, the theorem is as follows\footnote{\cite{Gas,Glau1,Glau2,Glau3} considered the more general case, where the values of $\rho$ and $m$ are different for different directions in $\mathbb{R}^d$. In this paper, we take common values of $\rho$ and $m$ for every direction, for simplicity.}.
\begin{theorem}
	Let $d$ and $m$ be positive integers.
	Let $\mathcal{D}$ be a hyper-rectangle like (\ref{eq:hypRect}).
	Let $f:\mathcal{D}\rightarrow \mathbb{R}$ be a function that has an analytic
	extension to $\mathcal{B}_{\mathcal{D},\rho}$ for some $\rho\in\mathbb{R}_{>1}$.
	Besides, assume that $\sup_{\vec{s}\in\mathcal{B}_{\mathcal{D},\rho}}|f(\vec{s})|\le B$ for some $B\in\mathbb{R}$.
	Moreover, suppose that we are given a real number $\hat{f}_{\vec{j}}$ for every $\vec{j}\in [m]_0^d$, and that there exists $\epsilon\in\mathbb{R}$ such that
	\begin{equation}
		\left|f\left(\vec{S}^{\mathcal{D},m}_{\vec{j}}\right)-\hat{f}_{\vec{j}}\right| \le \epsilon
	\end{equation}
	holds for every $\vec{j}\in [m]_0^d$.
	Then,
	\begin{equation}
		\max_{\vec{S}\in\mathcal{D}}|f(\vec{S})-\tilde{f}(\vec{S})|\le \epsilon^{\rm int}(\rho,d,m,B)+\Lambda_{d,m}\epsilon
	\end{equation}
	holds.
	Here, for every $\vec{S}\in\mathcal{D}$, $\tilde{f}(\vec{S})$ is defined as
	\begin{equation}
		\tilde{f}(\vec{S}) := \sum_{\vec{l}\in [m]_0^d} \tilde{a}_{\vec{l}} \widetilde{T}_{\mathcal{D},\vec{l}} \ (\vec{S}),
	\end{equation}
	with the coefficients $\tilde{a}_{\vec{l}}$ calculated by
	\begin{equation}
		\tilde{a}_{\vec{l}} := \frac{2^{\aleph\left(\vec{l}\right)}}{(m+1)^d}\sum_{\vec{j}\in [m]_0^d} \hat{f}_{\vec{j}}\widetilde{T}_{\mathcal{D},\vec{l}} \ \left(\vec{S}^{\mathcal{D},m}_{\vec{j}}\right)
	\end{equation}
	for every $\vec{l}\in[m]_0^d$, and
	\begin{equation}
		\epsilon^{\rm int}(\rho,d,m,B):=2^{\frac{d}{2}+1}\sqrt{d}B\rho^{-m}\left(1-\rho^{-2}\right)^{-\frac{d}{2}}.
	\end{equation}
	\label{th:ChebErrFunc}
\end{theorem}

\section{\label{sec:Berm}Bermudan option pricing}

\subsection{General formulation}

In this paper, we consider pricing a Bermudan option with $d\in\mathbb{N}$ underlying assets and $K\in\mathbb{N}$ exercise dates $t_1,...,t_K$, which satisfy $t_0<t_1<...<t_K$ with $t_0:=0$ being the present and $t_K:=T\in\mathbb{R}_+$ being the final maturity.
This is formulated as follows.
Under some filtered probability space $(\Omega,\mathcal{F},(\mathcal{F}_t)_{t\le0},\mathbb{P})$, consider the $\mathcal{S}$-valued Markov process $\vec{S}(t):=(S_1(t),...,S_d(t))^T$, where $\mathcal{S}$ is a subset of $\mathbb{R}^d$ equipped with its Borel $\sigma$-algebra inherited from $\mathbb{R}^d$, and $\vec{S}_0:=\vec{S}(0)$ is deterministic.
$\vec{S}(t)$ corresponds to the values of the underlying asset prices at time $t$, or transformations of them by some function (for example, the logarithms of the asset prices).
We are mainly interested in its values at $t_1,...,t_K$, that is, the discrete-time process $\vec{S}_k=(S_{1,k},...,S_{d,k}):=\vec{S}(t_k),k\in[K]_0$.
We hereafter denote an instance of this process, which is a $(K+1)$-tuple of elements of $\mathcal{S}$, as $\mathbf{S}=(\vec{S}_0,\vec{S}_1,...,\vec{S}_K)$.
Besides, suppose that we are given the function $f^{\rm pay}_k\in L^2(\mathcal{S},\rho_k)$ for every $k\in[K]$, where $\rho_k$ is the image probability measure on $\mathcal{S}$ induced by $\vec{S}_k$.
This corresponds to the payoff which arises by the exercise at $t_k$.
Although we assume that the risk-free rate is 0 for simplicity in this paper, we can consider that $f^{\rm pay}_k$ is the discounted payoff, that is, the product of the payoff and the discount factor.
Then, the price of the Bermudan option at $t_k$ with $\vec{S}_k=\vec{s}\in\mathcal{S}$ is given as
\begin{equation}
	V_k(\vec{s}):=\sup_{\tau\in\mathcal{T}_{k}} \mathbb{E}[f^{\rm pay}_\tau(\vec{S}_\tau)|\vec{S}_t=\vec{s}],
\end{equation}
where $\mathbb{E}[\cdot]$ denotes the (conditional or unconditional) expected value with respect to $\mathbb{P}$, and $\mathcal{T}_{k},k\in[K]$ is the set of all $\{k,...,K\}$-valued stopping times.
In particular, the present option price is
\begin{equation}
	V_0(\vec{s}):=\sup_{\tau\in\mathcal{T}} \mathbb{E}[f^{\rm pay}_\tau(\vec{S}_\tau)], \label{eq:V0}
\end{equation}
where $\mathcal{T}=\mathcal{T}_1$.

The problem to find $V_0$ can be written as a kind of dynamic programming, that is,
\begin{equation}
	V_k(\vec{s})=
	\begin{cases}
		f^{\rm pay}_K(\vec{s}) & ; \ k=K \\
		\max \{f^{\rm pay}_k(\vec{s}),Q_k(\vec{s})\} & ; \ k=1,...,K-1
	\end{cases}
\end{equation}
for every $\vec{s}\in\mathcal{S}$, and
\begin{equation}
	V_0 =\mathbb{E}[V_{1}(\vec{S}_{1})].
\end{equation}
Here, for every $k\in[K-1]$ and $\vec{s}\in\mathcal{S}$,
\begin{equation}
	Q_k(\vec{s}):=\mathbb{E}[V_{k+1}(\vec{S}_{k+1})|\vec{S}_k=\vec{s}].
\end{equation}
is called the continuation value.
This corresponds to the option price at $t_k$ in the case that the option has not been exercised at that time and that $\vec{S}_k=\vec{s}$.

Note that this problem can be seen as finding the optimal exercise date $\tau_{\rm op}\in\mathcal{T}$, which maximizes (\ref{eq:V0}).
This can be recursively determined as
\begin{eqnarray}
	\tau_K & = & K \nonumber \\
	\tau_k & = & k1_{f^{\rm pay}_k(\vec{S}_k)\ge Q_k(\vec{S}_k)} +  \tau_{k+1}1_{f^{\rm pay}_k(\vec{S}_k)< Q_k(\vec{S}_k)}, k\in[K-1]\nonumber \\
	&&
\end{eqnarray}
and $\tau_{\rm op}=\tau_1$.
Also note that
\begin{equation}
	Q_k(\vec{s})=\mathbb{E}[f^{\rm pay}_{\tau_{k+1}}(\vec{S}_{\tau_{k+1}})|\vec{S}_k=\vec{s}],
\end{equation}
for every $k\in[K-1]$, which means that $Q_k$ is the expected value of the payoff under the exercise strategy $\tau_{k+1}$.

\subsection{Least squares Monte Carlo \label{sec:LSM}}

We here explain LSM\cite{Longstaff}, as an algorithm for Bermudan option pricing.
This is one of the widely used methods in practical business, and the theoretical error bound on the price in the method has been investigated\cite{Clement,Glasserman2,Stentoft,Egloff,Gobet,Zanger1,Gerhold,Zanger2,Zanger3,Zanger4}. 

Omitting some technical details, we describe the outline of LSM as follows.
As a preparation, for every $k\in[K-1]$, we determine the set of functions $\mathcal{H}_k\subseteq L^2(\mathcal{S},\rho_k)$ for approximation of the continuation value $Q_k$. 
One common choice is $\mathcal{H}_k=\mathcal{R}_{d,m}$, the set of all real-coefficient polynomials on $\mathbb{R}^d$ of degree at most $m\in \mathbb{N}$, and we hereafter consider this.
Next, we generate $N_{\rm samp}\in \mathbb{N}_{\ge2}$ sample paths of underlying asset prices, which are denoted as $\mathbf{S}_i=(\vec{S}_0,\vec{S}^{(i)}_1,...,\vec{S}^{(i)}_K)$, $i\in[N_{\rm samp}]$.
Then, we determine the stopping time, which approximate the optimal one $\tau_{\rm op}$, by the following procedure.
First, we set $\widehat{\tau}_{K,i}=K$ for every $i\in[N_{\rm samp}]$.
For $k\in[K-1]$, given $\widehat{\tau}_{k+1,i}$ for every $i\in[N_{\rm samp}]$, we determine the approximate continuation value $\widehat{Q}_k$ as the element $g_k$ in $\mathcal{R}_{d,m}$, which minimizes
\begin{equation}
	\frac{1}{N_{\rm samp}}\sum_{i=1}^{N_{\rm samp}} \left(g_k(\vec{S}_k^{(i)})-f^{\rm pay}_{\widehat{\tau}_{k+1,i}}(\vec{S}^{(i)}_{\widehat{\tau}_{k+1,i}})\right)^2, \label{eq:sqErr}
\end{equation}
or, in other words, best fits to the realized payoffs under the stopping time $\widehat{\tau}_{k+1,i}$ on the sample paths.
It is guaranteed by statistical leaning theory that fitting to the sample values of the payoff, which distribute around the continuation value, yields the approximation of the continuation value~\cite{Egloff,Zanger1,Zanger2,Zanger3,Zanger4}.
Then, we set
\begin{equation}
\widehat{\tau}_{k,i}=
\begin{cases}
	k & ; \ {\rm if} \ \widehat{Q}_k(\vec{S}^{(i)}_k)\le f^{\rm pay}_{k}(\vec{S}^{(i)}_k) \\
	\widehat{\tau}_{k+1,i} & ; \ {\rm otherwise}
\end{cases}
\end{equation}
for every $i\in[N_{\rm samp}]$.
By repeating this until we reach $k=1$, we get $\widehat{\tau}_{1,i}$, and finally 
\begin{equation}
\widehat{V}_0:=\frac{1}{N_{\rm samp}}\sum_{i=1}^{N_{\rm samp}}f^{\rm pay}_{\widehat{\tau}_{1,i}}(\vec{S}^{(i)}_{\widehat{\tau}_{1,i}}) \nonumber
\end{equation}
as an approximation of $V_0$.

Let us make some comments on the procedure.
First, note that it is assumed that we can generate sample paths $\mathbf{S}_i$.
In the usual situation where the stochastic differential equation (SDE) for $\vec{S}(t)$ is given, we can use some method for numerical simulation of SDEs such as Euler-Maruyama method.
Second, we mention how to find $g_k$ minimizing (\ref{eq:sqErr}). 
Note that this is just least-squares linear regression, since $\mathcal{R}_{d,m}$ is a vector space.
Therefore, we can solve this by various methods, for example, solving the normal equation of linear regression, some numerical optimization, and so on.

Then, let us mention the relationship between the error and the sample number in LSM.
According to \cite{Zanger4}, under some technical assumptions, taking appropriately large $m$, we obtain the error bound on the option price which scales as
\begin{equation}
	\mathbb{E}_{\rm samp}[|\widehat{V}_0-V_0|] =\widetilde{O}\left(\left(N_{\rm samp}\right)^{-\frac{n(p-2)}{2n(p+2)+d(p-2)}}\right). \label{eq:LSMErr}
\end{equation}
Here, $\mathbb{E}_{\rm samp}[\cdot]$ denotes the expectation with respect to randomness of samples, and $n$ and $p$ are the quantities which characterize smoothness of $Q_1(\vec{S}),...,Q_{K-1}(\vec{S})$ and boundedness of the norms of $f^{\rm pay}_1(\vec{S}),...,f^{\rm pay}_K(\vec{S})$, respectively (see \cite{Zanger4} for more details).
For larger $p$ and $n$, the RHS of (\ref{eq:LSMErr}) decreases faster against the increase of $N_{\rm samp}$.
In the limit of $n,p\rightarrow \infty$, which means that $Q_k$'s are highly smooth and the norms of $f^{\rm pay}_k$'s are well-bounded, the RHS of (\ref{eq:LSMErr}) becomes $\widetilde{O}(N_{\rm samp}^{-1/2})$, which coincides with the well-known error decay rate in Monte Carlo integration.
Conversely, in this limit, it is sufficient to take $\widetilde{O}(\epsilon^{-2})$ samples in order to achieve the error tolerance $\epsilon$.

\section{\label{sec:QAE}Quantum amplitude estimation and quantum algorithm for Monte Carlo integration} 

\subsection{Quantum amplitude estimation (QAE)}

We here briefly review QAE.
Consider the system consisting of a quantum register $R_1$ and a qubit $R_2$.
Suppose that we are given the oracle $A$, which transforms $\ket{0}\ket{0}$, the state in which all qubits in $R_1$ and $R_2$ are set to $\ket{0}$, into
\begin{equation}
	A\ket{0}\ket{0}=\sqrt{a}\ket{\psi_1}\ket{1}+\sqrt{1-a}\ket{\psi_0}\ket{0}=:\ket{\Psi} \label{eq:oracleQAE}
\end{equation}
with some $a\in(0,1)$.
Here, the first and second kets correspond to $R_1$ and $R_2$ respectively, and $\ket{\psi_0},\ket{\psi_1}$ are some quantum states.
Then, our goal is estimating $a$, which is the probability to obtain 1 in $R_2$ when we measure $\ket{\Psi}$, with the error tolerance $\epsilon$.
There exist some algorithms which output such an estimation with $O(\epsilon^{-1})$ calls to $A$ and its inverse $A^\dagger$ in total\cite{Brassard,Suzuki,Aaronson,Grinko,Nakaji}.
Although these QAE algorithms output a number close to $a$ not with certainty but with some probability, we can enhance the success probability by running QAE many times and take the median of the outputs\cite{Montanaro,Jerrum}. 
Let us define the $(N_{\rm QAE},N_{\rm rep})$-QAE as the method for estimating $a$ which runs $N_{\rm rep}$ rounds of QAE and makes $N_{\rm QAE}$ calls to $A$ and $A^\dagger$ in total in each round.
Obviously, in the $(N_{\rm QAE},N_{\rm rep})$-QAE, $A$ and $A^\dagger$ are called $N_{\rm QAE}N_{\rm rep}$ times in total.
Then, combining Theorem 12 in \cite{Brassard} and Lemma 6.1 in \cite{Jerrum}, we obtain the following theorem.

\begin{theorem}
	Suppose that we are given the accesses to $A$ in (\ref{eq:oracleQAE}) and its inverse $A^\dagger$.
	Then, for any $\gamma\in(0,1)$ and $\epsilon\in(0,0.1)$, a $(N_{\rm QAE},N_{\rm rep})$-QAE, where the positive integers $N_{\rm QAE}$ and $N_{\rm rep}$ satisfy
	\begin{equation}
		N_{\rm QAE} \ge \frac{7}{\epsilon}  \label{eq:NQAE}
	\end{equation}
	and
	\begin{equation}
		N_{\rm rep} \ge 12\lceil\log(\gamma^{-1})\rceil + 1
	\end{equation}
	respectively, outputs $\tilde{a}\in\mathbb{R}$ such that
	\begin{equation}
		|\tilde{a}-a|\le \frac{7}{N_{\rm QAE}}\le \epsilon,
	\end{equation}
	where $a$ is defined as (\ref{eq:oracleQAE}),
	with probability higher than $1-\gamma$. \label{th:QAE}
\end{theorem}

Here, (\ref{eq:NQAE}) is derived from the inequality in Theorem 12 in \cite{Brassard} with $k=1$, that is,
\begin{equation}
	|\tilde{a}-a|\le \frac{2\pi\sqrt{a(1-a)}}{M} + \frac{\pi^2}{M^2}, \label{eq:QAEErrOri}
\end{equation}
where 
$M=N_{\rm QAE}/2$ under the current definition.
Using $\sqrt{a(1-a)}\le \frac{1}{2}$, we see that (\ref{eq:QAEErrOri}) implies $|\tilde{a}-a|\le 7/N_{\rm QAE}$ for $N_{\rm QAE}\ge 70$, which follows from (\ref{eq:NQAE}) and $0<\epsilon<0.1$.
In summary, if $N_{\rm QAE}$ satisfies (\ref{eq:NQAE}) for $\epsilon\in(0,0.1)$, the error in QAE is suppressed to at most $\epsilon$ with high probability.
Hereafter, we say that a $(N_{\rm QAE},N_{\rm rep})$-QAE succeeded if it output $\tilde{a}$ such that $|\tilde{a}-a|\le 7/N_{\rm QAE}$.

\subsection{Quantum algorithm for Monte Carlo integration}

One application of QAE is the algorithm for Monte Carlo integration, that is, the method to calculate expected values.
Suppose that we want to calculate $\mathbb{E}[F(\vec{X})]$, the expected value of $F(\vec{X})$, where $\vec{X}$ is some real vector-valued stochastic variable and $F$ is a real-valued function acting on $\vec{X}$.
We also assume that the range of $F$ is in $[0,1]$, and, if not, we make this hold by adding and/or multiplying some constants to $F$.
Furthermore, suppose that we can use the following oracles $O_{\vec{X}}$ and $O_{F}$.
$O_{\vec{X}}$ is the oracle to generate the state in which the distribution of $\vec{X}$ is encoded.
That is, $O_{\vec{X}}$ operates on a quantum register and transform the state with all qubits set to $\ket{0}$ into
\begin{equation}
	O_{\vec{X}}\ket{0}=\sum_{i=1}^{N_{\vec{X}}} \sqrt{p_i}\ket{\vec{x}_i},
\end{equation}
where $\vec{x}_1,...,\vec{x}_{N_{\vec{X}}}$ are $N_{\vec{X}}\in\mathbb{N}$ possible values of $\vec{X}$ and $p_i,i\in[N_{\vec{X}}]$ is the probability that $\vec{X}=\vec{x}_i$.
Here, we assume that the set of all the values that $\vec{X}$ can take is finite.
If $\vec{X}$ is continuous, we need some discretization.
How to create states corresponding to widely used distributions such as normal distribution has been investigated\cite{Grover,Kaneko}.
The second oracle $O_{F}$ operates on a two-register system, and, using the first register as the input $\vec{x}$, outputs $F(\vec{x})$ into the second register.
That is, for any $\vec{x}$ in the domain of $F$,
\begin{equation}
	O_F\ket{\vec{x}}\ket{0}=\ket{\vec{x}}\ket{F(\vec{x})}.
\end{equation}
By these oracles, the following computation is possible.
Preparing two registers $R_1,R_2$ and a qubit $R_3$, and initializing all of them to $\ket{0}$, we perform
\begin{eqnarray}
&&	\ket{0}\ket{0}\ket{0} \nonumber \\
&\rightarrow& \sum_{i=1}^{N_{\vec{X}}} \sqrt{p_i}\ket{\vec{x}_i}\ket{0}\ket{0} \nonumber \\
&\rightarrow& \sum_{i=1}^{N_{\vec{X}}} \sqrt{p_i}\ket{\vec{x}_i}\ket{F(\vec{x}_i)}\ket{0} \nonumber \\
&\rightarrow& \sum_{i=1}^{N_{\vec{X}}} \sqrt{p_i}\ket{\vec{x}_i}\ket{F(\vec{x}_i)}\left(\sqrt{F(\vec{x}_i)}\ket{1}+\sqrt{1-F(\vec{x}_i)}\ket{0}\right),
\nonumber \\
&& \label{eq:QMCTrans}
\end{eqnarray}
where the first, second and third kets correspond to $R_1$, $R_2$ and $R_3$, respectively.
We use $O_{\vec{X}}$ and $O_F$ at the first and second arrows, respectively.
The transformation at the third arrow is done by arithmetic circuits\cite{Haner} and controlled rotation gates.
Note that the probability to obtain 1 in $R_3$ when we measure the final state in (\ref{eq:QMCTrans}) is $\sum_{i=1}^{N_{\vec{X}}}p_iF(\vec{x}_i)$, that is, $\mathbb{E}[F(\vec{X})]$.
Therefore, using the whole operation in (\ref{eq:QMCTrans}) as the oracle $A$, we can estimate $\mathbb{E}[F(\vec{X})]$ by QAE.

\section{\label{sec:NewAlgo}Bermudan option pricing by Chebyshev interpolation and QAE}


Now, let us present the method for Bermudan option pricing by Chebyshev interpolation and QAE.

\subsection{Assumptions}

We begin with making some assumptions necessary to execute the proposed method.
The first one is as follows.

\begin{assumption}
	We are given the access to the oracle $O_{{\rm step},k}$, which generates the state corresponding to the probability distribution of $\vec{S}_{k+1}$ conditional on $\vec{S}_{k}$.
	That is, for every $k\in[K-1]_0$ and $\vec{S}\in \mathcal{S}$,
	\begin{equation}
		O_{{\rm step},k}:\ket{\vec{S}}\ket{0}\mapsto \sum_{\vec{s}\in \widetilde{\mathcal{S}}_{k+1}(\vec{S})}\sqrt{p_{k+1}(\vec{s};\vec{S})}\ket{\vec{S}}\ket{\vec{s}}, \label{eq:oracleStep}
	\end{equation}
	where $\widetilde{\mathcal{S}}_{k+1}(\vec{S})$ is the set of possible values of $\vec{S}_{k+1}$ under the condition that $\vec{S}_{k}=\vec{S}$, and
	\begin{equation}
	p_{k+1}(\vec{s};\vec{S}):=\mathbb{P}\left(\vec{S}_{k+1}=\vec{s} \ \middle| \ \vec{S}_k=\vec{S}\right).
	\end{equation}
	\label{ass:EvolOra}
\end{assumption}

We here make comments on how to implement $O_{{\rm step},k}$.
As mentioned in Section \ref{sec:Berm}, usually, following some SDE and some numerical method such as Euler-Maruyama, we can generate random sample values of $\vec{S}_{k+1}$ with the given value of $\vec{S}_k$ as the initial condition.
Implementations of such a calculation on quantum circuits have been discussed in the previous papers\cite{Rebentrost,Stamatopoulos,Kaneko}.
That is, we can prepare the states corresponding to some (discretely approximated) random variables (e.g. standard normal) on the other registers, and, using them at discretized time steps, generate the path of $\vec{S}(t)$ from $t_k$ to $t_{k+1}$.
This yields the state like (\ref{eq:oracleStep}).
We should also note that, in Assumption \ref{ass:EvolOra}, it is assumed that $\vec{S}_{k+1}$ can take only a finite number of values for the fixed $\vec{S}_k$.
This is not the case in the most models of $\vec{S}(t)$, in which it takes continuous values.
However, under the aforementioned implementations for time evolution of $\vec{S}(t)$, in which both time and random variables are discretely approximated, the number of the possible values of $\vec{S}_k$ necessarily becomes finite.

Hereafter, we are mainly interested in the number of calls to $O_{{\rm step},k}$ in calculating the option price as a measure of complexity, since calculation for time evolution of underlying asset prices is typically the most time-consuming part in option pricing.

The second assumption is as follows.
Here, $\mathcal{I}_{\mathcal{A}}$ denotes the set of all real-valued functions on a given subset $\mathcal{A}\subseteq\mathbb{R}^d$.
\begin{assumption}
	For every $k\in[K-1]$, we are given the following
	\begin{itemize}
		\item the hyper-rectangle $\mathcal{D}_k:=[L_{1,k},U_{1,k}]\times...\times[L_{d,k},U_{d,k}]\subseteq \mathcal{S}$, with $L_{1,k},...,L_{d,k},U_{1,k},...,U_{d,k}\in\mathbb{R}$ satisfying $L_{1,k}<U_{1,k},...,L_{d,k}<U_{d,k}$,
		
		\item $V^{\rm OB}_k\in\mathcal{I}_{\mathcal{S}\setminus \mathcal{D}_k}$
		
	\end{itemize}
	such that the following (i) and (ii) are satisfied.
	\begin{enumerate}
		\renewcommand{\labelenumi}{(\roman{enumi})}
		
		
		\item There exists $\epsilon^{\rm OB}_k\in\mathbb{R}_+$ such that either
		\begin{equation}
			|V^{\rm OB}_k(\vec{s})-V_k(\vec{s})| < \epsilon^{\rm OB}_k \label{eq:ass2Cond1}
		\end{equation}
		or
		\begin{equation}
			|\mathbf{F}_{k}[V_k](\vec{s})-V_k(\vec{s})| < \epsilon^{\rm OB}_k \label{eq:ass2Cond2}
		\end{equation}
		is satisfied for any $\vec{s}\in\mathcal{S}\setminus\mathcal{D}_k$.
		Here, $\mathbf{F}_{k}[\cdot]$ is the `flat extrapolation operator' defined as
		\begin{eqnarray}
			\mathbf{F}_{k}[F](\vec{s}) &:=& F(b_k(\vec{s})) \\
			b_k(\vec{s}) &:=&\left(\min\{U_{1,k},\max\{L_{1,k},s_1\}\},..., \right.\nonumber \\
			&&\qquad\qquad \left.\min\{U_{d,k},\max\{L_{d,k},s_d\}\}\right)^T 
		\end{eqnarray}
		for any $F\in\mathcal{I}_{\mathcal{D}_k}$ and $\vec{s}=(s_1,...,s_d)^T\in\mathcal{S}$.
		
		\item If, for some $G\in\mathcal{I}_{\mathcal{D}_k}$, we have the access to the oracle $O_G$ such that
		\begin{equation}
			O_G\ket{\vec{s}}\ket{0}=\ket{\vec{s}}\ket{G(\vec{s})}
		\end{equation}
		for any $\vec{s}\in\mathcal{D}_k$, we also have the access to the oracle $\widetilde{O}_G$, which acts as
		\begin{equation}
			\widetilde{O}_G\ket{\vec{s}}\ket{0}=\ket{\vec{s}}\ket{\mathbf{G}_{k}[G](\vec{s})}.
		\end{equation}
		Here, $\mathbf{G}_{k}[\cdot]$ is defined as
		\begin{equation}
			\mathbf{G}_{k}[H](\vec{s}) :=
			\begin{cases}
				V^{\rm OB}_k(\vec{s}) & ; \ {\rm if} \ \vec{s}\in\mathcal{A}_k \\
				\mathbf{F}_k[H](\vec{s}) & ; \ {\rm otherwise}
			\end{cases}
		\end{equation}
		for any $H\in\mathcal{I}_{\mathcal{D}_k}$ and $\vec{s}\in\mathcal{S}$, where $\mathcal{A}_k$ is a subset of $\mathcal{S}\setminus \mathcal{D}_k$ such that (\ref{eq:ass2Cond1}) and (\ref{eq:ass2Cond2}) hold for any $\vec{s}\in\mathcal{A}_k$ and any $\vec{s}\in(\mathcal{S}\setminus \mathcal{D}_k)\setminus\mathcal{A}_k$, respectively.
		
	\end{enumerate}
	%
	%
	%
	
	\label{ass:Extrap}
\end{assumption}
\noindent We also define $\mathbf{G}_{K}[H](\vec{s}) := H(\vec{s})$ for any $H\in\mathcal{I}_{\mathcal{S}}$ and $\vec{s}\in\mathcal{S}$

Roughly speaking, this assumption means that, when some of underlying asset prices are extremely large or small, we can approximate the option value $V_k$ by some known and easily computable function $V^{\rm OB}_k$ or the flat extrapolation of $V_k$ from moderate underlying asset prices.  
Postponing explanation on why this assumption is necessary to Section \ref{sec:NewAlgoDet}, we here see that it is actually satisfied in some typical settings in option pricing.
For example, let us consider a basket put option, whose payoff function is $f^{\rm pay}_k((s_1,...,s_d)^T)=\max\{\kappa-s_1-...-s_d,0\}$ with some $\kappa\in\mathbb{R}$ for every $k\in[K]$, under some model in which $S_1(t),...,S_d(t)$ are unbounded from above but bounded from below, say, by 0, as the Black-Scholes model.
Then, in each of the following situations, (\ref{eq:ass2Cond1}) or (\ref{eq:ass2Cond2}) holds.
\begin{itemize}
	\item If some of $S_{1,k},...,S_{d,k}$ are extremely large, the option is far out-of-money, and therefore its price is almost 0.
	\item If some of $S_{1,k},...,S_{d,k}$ are smaller than the sufficiently small thresholds $L_{1,k},...,L_{d,k}\in\mathbb{R}_+$ respectively, but the others are not, setting the former to the thresholds hardly affects the option price.
	\item If all of $S_{1,k},...,S_{d,k}$ are sufficiently close to 0, the option is exercised at $t_k$, and therefore $V_k(\vec{S}_k)=f^{\rm pay}_k(\vec{S}_k)$.
\end{itemize}

Thirdly, we make the following assumption, which is necessary for bounding the interpolation error in the proposed method.
\begin{assumption}
	For every $k\in[K-1]$, $Q_k(\vec{S})$ has an analytic extension to $\mathcal{B}_{\mathcal{D}_k,\rho_k}$, where $\mathcal{D}_k$ is given in Assumption \ref{ass:Extrap} and $\rho_k$ is some real number greater than 1, and
	\begin{equation}
		\sup_{\vec{s}\in \mathcal{B}_{\mathcal{D}_k,\rho_k}} |Q_k(\vec{S})| \le B_k
	\end{equation}
	holds, where $B_k$ is some positive real number.
	\label{ass:QkAnal}
\end{assumption}

\subsection{The proposed method \label{sec:NewAlgoDet}}

Under these assumptions, we can construct the procedure for Bermudan option pricing based on QAE and Chebyshev interpolation.
This is also a backward calculation similarly to LSM; we sequentially calculate the approximate continuation value $\widetilde{Q}_k$ and option price $\widetilde{V}_k$ at $t_k$, going from the final maturity to the present.
Roughly, the outline is as follows.
As preparation, for every $k\in [K-1]$, we set $m_k\in\mathbb{N}$, the degree of Chebyshev polynomials used for the approximation, and the hyper-rectangle $\mathcal{D}_k=[L_{1,k},U_{1,k}]\times\cdots\times[L_{d,k},U_{d,k}]\subseteq \mathcal{S}$.   
We begin the iterative calculation by setting $\widetilde{V}_K(\vec{S}):=f^{\rm pay}_K(\vec{S})$ for every $\vec{S}\in\mathcal{S}$.
Then, for $k\in[K-1]$, given $\widetilde{V}_{k+1}$, we estimate the expected value of $\widetilde{V}_{k+1}(\vec{S}_{k+1})$ under the condition that $\vec{S}_k=\vec{S}^{\mathcal{D}_k,m_k}_{\vec{j}}$ for every Chebyshev node $\vec{S}^{\mathcal{D}_k,m_k}_{\vec{j}}$ by QAE, and denote the estimation as $\widehat{Q}_{k,\vec{j}}^{\rm QAE}$.
Using these, we construct $\widetilde{Q}_k$, the Chebyshev interpolation of the approximate continuation value, and set $\widetilde{V}_k(\vec{S})=\max\{\widetilde{Q}_k(\vec{S}),f^{\rm pay}_k(\vec{S})\}$ for every $\vec{S}\in\mathcal{S}$.
We repeat these steps until we reach $k=1$.
Finally, we estimate the expected value of $\widetilde{V}_1(\vec{S}_1)$ by QAE again, and let the result be an approximation of $V_0$.

The fully detailed procedure is shown in Algorithm \ref{alg:new}.

\begin{algorithm}[H]
	\caption{The method for Bermudan option pricing based on Chebyshev interpolation and QAE}
	\label{alg:new}
	\begin{algorithmic}[1]
		\REQUIRE{\ \\
			\begin{itemize}
				\item $m_k\in\mathbb{N}$ for every $k\in[K-1]$, the degree of Chebyshev polynomials.  \\
				\item $N^{\rm QAE}_{k}\in\mathbb{N}$ for every $k\in[K-1]_0$, the iteration number in each run of QAE in calculating $\widehat{Q}^{\rm QAE}_{k,\vec{j}}$ or $\widetilde{V}_0$.\\
				\item $N^{\rm rep}_{k}\in\mathbb{N}$ for every $k\in[K-1]_0$, the number of rounds of QAE in calculating $\widehat{Q}^{\rm QAE}_{k,\vec{j}}$ or $\widetilde{V}_0$.\\
				\item $\widetilde{V}^{\rm max}_{k}\in\mathbb{R}_+$ for every $k\in[K]$, the upper bound of $\{|\mathbf{G}_{k}[\hat{V}_{k}](\vec{s})| \ | \ \vec{s}\in\mathcal{S}\}$.
				\item  $\mathcal{D}_k=[L_{1,k},U_{1,k}]\times\cdots\times[L_{d,k},U_{d,k}]\subseteq \mathcal{S}$ for every $k\in[K-1]$, the hyper-rectangle for Chebyshev interpolation. Here, $L_{1,k},...,L_{d,k},U_{1,k},...,U_{d,k}\in\mathcal{S}$ satisfy $L_{1,k}<U_{1,k},...,L_{d,k}<U_{d,k}$.
			\end{itemize}
		}
		\STATE Set $\widetilde{V}_K(\vec{S}):=f^{\rm pay}_K(\vec{S})$ for every $\vec{S}\in\mathcal{S}$.	
		
		\FOR{$k = K-1$ to $1$}
		\FORALL{$\vec{j}\in \mathcal{J}_k:=[m_k]^d_0$}
		
		\STATE Using $(N^{\rm QAE}_{k},N^{\rm rep}_{k})$-QAE, obtain an estimation $\widetilde{P}_{k,\vec{j}}$ of the probability $P_{k,\vec{j}}$ to observe $1$ on the last qubit in measuring $\ket{\Psi_{k,\vec{j}}}$ in (\ref{eq:Psik}), and let $(2\widetilde{P}_{k,\vec{j}}-1)\widetilde{V}^{\rm max}_{k}$ be $\widehat{Q}^{\rm QAE}_{k,\vec{j}}$.

		\ENDFOR
		
		\STATE
		Set 
		\begin{equation}
			\widetilde{Q}_k(\vec{S}):=\sum_{\vec{l}\in\mathcal{J}_k} \tilde{a}_{k,\vec{l}} \ \widetilde{T}_{\mathcal{D}_k,\vec{l}} \ (\vec{S})
		\end{equation}
		for every $\vec{S}\in\mathcal{D}_k$, with $\tilde{a}_{k,\vec{l}}$ calculated as
		\begin{equation}
			a_{k,\vec{l}}:=	\frac{2^{\aleph\left(\vec{l}\right)}}{(m_k+1)^d}\sum_{\vec{j}\in \mathcal{J}_k} \widehat{Q}^{\rm QAE}_{k,\vec{j}}\widetilde{T}_{\mathcal{D}_k,\vec{l}} \ \left(\vec{S}^{\mathcal{D}_k,m_k}_{\vec{j}}\right)
		\end{equation}
		for $\vec{l}\in\mathcal{J}_k$.
		
		\STATE Set $\widetilde{V}_{k}(\vec{S}):=\max\left\{f^{\rm pay}_k(\vec{S}),\widetilde{Q}_k(\vec{S})\right\}$ for any $\vec{S}\in\mathcal{D}_k$.
		
		\ENDFOR
		
		\STATE Using $(N^{\rm QAE}_{0},N^{\rm rep}_{0})$-QAE, obtain an estimation $\widetilde{P}_0$ of the probability $P_0$ to observe $1$ on the last qubit in measuring $\ket{\Psi_{0}}$ in (\ref{eq:Psi0}), and output $(2\widetilde{P}_0-1)\widetilde{V}^{\rm max}_{1}$ as $\widetilde{V}_0$.
	\end{algorithmic}
\end{algorithm}

Some additional explanations should be made.
The first one is on $\ket{\Psi_{k,\vec{j}}}$ in Step 4.
For every $k\in[K-1]$ and $\vec{j}\in\mathcal{J}_k$, given the approximation $\widetilde{V}_{k+1}\in\mathcal{I}_{\mathcal{D}_{k+1}}$ of $V_{k+1}$, we generate the state $\ket{\Psi_{k,\vec{l}}}$ on the appropriate multi-register system with the last one being single-qubit, by the following operation:
\begin{eqnarray}
	&&\ket{0}\ket{0}\ket{0}\ket{0} \nonumber \\
	&\rightarrow& \Ket{\vec{S}^{\mathcal{D}_k,m_k}_{\vec{j}}}\ket{0}\ket{0}\ket{0} \nonumber \\
	&\rightarrow& \Ket{\vec{S}^{\mathcal{D}_k,m_k}_{\vec{j}}}\sum_{\vec{s}\in\widetilde{\mathcal{S}}_{k+1}\left(\vec{S}^{\mathcal{D}_k,m_k}_{\vec{j}}\right)} \sqrt{p_{k+1}\left(\vec{s};\vec{S}^{\mathcal{D}_k,m_k}_{\vec{j}}\right)}\ket{\vec{s}}\ket{0}\ket{0} \nonumber \\
	&\rightarrow& \Ket{\vec{S}^{\mathcal{D}_k,m_k}_{\vec{j}}}\sum_{\vec{s}\in\widetilde{\mathcal{S}}_{k+1}\left(\vec{S}^{\mathcal{D}_k,m_k}_{\vec{j}}\right)} \sqrt{p_{k+1}\left(\vec{s};\vec{S}^{\mathcal{D}_k,m_k}_{\vec{j}}\right)}\ket{\vec{s}}\ket{\mathbf{G}_{k+1}[\widetilde{V}_{k+1}](\vec{s})}\ket{0} \nonumber \\
	&\rightarrow& \Ket{\vec{S}^{\mathcal{D}_k,m_k}_{\vec{j}}}\sum_{\vec{s}\in\widetilde{\mathcal{S}}_{k+1}\left(\vec{S}^{\mathcal{D}_k,m_k}_{\vec{j}}\right)} \sqrt{p_{k+1}\left(\vec{s};\vec{S}^{\mathcal{D}_k,m_k}_{\vec{j}}\right)}\ket{\vec{s}}\ket{\mathbf{G}_{k+1}[\widetilde{V}_{k+1}](\vec{s})}
	\nonumber\\	&&\qquad\qquad\qquad\qquad\qquad\otimes
	\left(\sqrt{\frac{1}{2}+\frac{\mathbf{G}_{k+1}[\widetilde{V}_{k+1}](\vec{s})}{2\widetilde{V}_{k+1}^{\rm max}}}\ket{1}\right. \nonumber \\
	&&\qquad\qquad\qquad\qquad\qquad\qquad\left.+\sqrt{\frac{1}{2}-\frac{\mathbf{G}_{k+1}[\widetilde{V}_{k+1}](\vec{s})}{2\widetilde{V}_{k+1}^{\rm max}}}\ket{0}\right) \nonumber \\
	&=:& \ket{\Psi_{k,\vec{j}}}, \label{eq:Psik}
\end{eqnarray}
where $O_{{\rm step},k}$ in Assumption \ref{ass:EvolOra} and $\widetilde{O}_{\widetilde{V}_{k+1}}$ in Assumption \ref{ass:Extrap} are used at the second and third arrows, respectively.
Note that the probability to obtain $1$ on the last qubit in measuring $\ket{\Psi_{k,\vec{l}}}$ is
\begin{equation}
P_{k,\vec{j}}=\frac{1}{2}+\frac{\widehat{Q}_k\left(\vec{S}^{\mathcal{D}_k,m_k}_{\vec{j}}\right)}{2\widetilde{V}_{k+1}^{\rm max}},
 \label{eq:Pkj}
\end{equation}
where
\begin{eqnarray}
	\widehat{Q}_k\left(\vec{S}\right)&:=&\mathbb{E}\left[\mathbf{G}_{k+1}[\widetilde{V}_{k+1}](\vec{S}_{k+1}) \ \middle| \ \vec{S}_k=\vec{S} \right] \nonumber \\
	&=&\sum_{\vec{s}\in\widetilde{\mathcal{S}}_{k+1}\left(\vec{S}\right)} p_{k+1}\left(\vec{s};\vec{S}\right)\mathbf{G}_{k+1}[\widetilde{V}_{k+1}](\vec{s}). \label{eq:Qhat}
\end{eqnarray}
Therefore, as long as $\mathbf{G}_{k+1}[\widetilde{V}_{k+1}]$ is close to $V_{k+1}$, $(2P_{k,\vec{j}}-1)\widetilde{V}^{\rm max}_{k+1}=\widehat{Q}_k\left(\vec{S}^{\mathcal{D}_k,m_k}_{\vec{j}}\right)$ 
is close to $Q_k\left(\vec{S}^{\mathcal{D}_k,m_k}_{\vec{j}}\right)$.
This is why we can obtain approximations of the continuation values at Chebyshev nodes by Step 4, with the errors from QAEs being also small.

Second, let us explain the state $\ket{\Psi_0}$ in Step 9.
Given $\widetilde{V}_1$, we can generate $\ket{\Psi_0}$ similarly to $\ket{\Psi_{k,\vec{j}}}$ as
\begin{eqnarray}
	&&\ket{0}\ket{0}\ket{0}\ket{0} \nonumber \\
	&\rightarrow&\ket{\vec{S}_0}\ket{0}\ket{0}\ket{0} \nonumber \\
	&\rightarrow& \ket{\vec{S}_0}\sum_{\vec{s}\in\widetilde{\mathcal{S}}_{1}(\vec{S}_0)} \sqrt{p_{1}(\vec{s};\vec{S}_0)}\ket{\vec{s}}\ket{0}\ket{0} \nonumber \\
	&\rightarrow&\ket{\vec{S}_0}\sum_{\vec{s}\in\widetilde{\mathcal{S}}_{1}(\vec{S}_0)} \sqrt{p_{1}(\vec{s};\vec{S}_0)}\ket{\vec{s}}\ket{\mathbf{G}_{1}[\widetilde{V}_{1}](\vec{s})}\ket{0} \nonumber \\
	&\rightarrow& \ket{\vec{S}_0}\sum_{\vec{s}\in\widetilde{\mathcal{S}}_{1}(\vec{S}_0)} \sqrt{p_{1}(\vec{s};\vec{S}_0)}\ket{\vec{s}}\ket{\mathbf{G}_{1}[\widetilde{V}_{1}](\vec{s})}
	\nonumber\\	&&\qquad\qquad\otimes
	\left(\sqrt{\frac{1}{2}+\frac{\mathbf{G}_{1}[\widetilde{V}_{1}](\vec{s})}{2\widetilde{V}_{1}^{\rm max}}}\ket{1}+\sqrt{\frac{1}{2}-\frac{\mathbf{G}_{1}[\widetilde{V}_{1}](\vec{s})}{2\widetilde{V}_{1}^{\rm max}}}\ket{0}\right) \nonumber \\
	&=:& \ket{\Psi_0}, \label{eq:Psi0}
\end{eqnarray}
where the last ket corresponds to a single-qubit register. 
Since the probability $P_0$ to obtain $1$ on the last qubit in measuring $\ket{\Psi_0}$ satisfies
\begin{equation}
	(2P_0-1)\widetilde{V}^{\rm max}_{1}=\widehat{V}_0,
\end{equation}
where
\begin{equation}
	\widehat{V}_0:=\mathbb{E}\left[\mathbf{G}_{1}[\widetilde{V}_{1}](\vec{S}_{1})\right]=\sum_{\vec{s}\in\widetilde{\mathcal{S}}_{1}(\vec{S}_0)} p_{1}(\vec{s};\vec{S}_0)\mathbf{G}_{1}[\widetilde{V}_{1}](\vec{s}),
\end{equation}
we can obtain an approximation of $V_0$ by Step 9, as long as $\mathbf{G}_{1}[\widetilde{V}_{1}]$ is close to $V_1$ and the QAE error is small.


Lastly, let us comment on the reason why Assumption \ref{ass:Extrap} is necessary.
This is because we have to handle underlying asset prices out of $\mathcal{D}_{k+1}$ in Step 4 and 9, or, more specifically, in generating $\ket{\Psi_{k,\vec{j}}}$ and $\ket{\Psi_0}$.
In fact, when we generate $\ket{\Psi_{k,\vec{j}}}$, $\vec{S}_{k+1}$ can be out of $\mathcal{D}_{k+1}$ with some probability.
In particular, when $\ket{\Psi_{k,\vec{j}}}$ corresponds to a Chebyshev node $\vec{S}^{\mathcal{D}_k,m_k}_{\vec{j}}$ close to the boundary of $\mathcal{D}_{k}$, or, in other words, the condition that $\vec{S}_k$ is close to the boundary of $\mathcal{D}_{k}$ is imposed, such a probability becomes non-negligible.

\subsection{Evaluation of the error}

Then, let us consider the error on the present option price in the proposed method.
First, we have the following theorem.

\begin{theorem}
	Under Assumptions \ref{ass:EvolOra} to \ref{ass:QkAnal}, consider Algorithm \ref{alg:new}.
	Suppose that, for every $k\in[K-1]$ and $\vec{j}\in \mathcal{J}_k$,
	\begin{equation}
		\left|\widehat{Q}_k\left(\vec{S}^{\mathcal{D}_k,m_k}_{\vec{j}}\right)-\widehat{Q}^{\rm QAE}_{k,\vec{j}}\right|\le \epsilon^{\rm QAE}_k \label{eq:assQAEErr}
	\end{equation}
	is satisfied, where $\epsilon^{\rm QAE}_k$ is some positive real number.
	Moreover, suppose that
	\begin{equation}
		\left|\widehat{V}_0-\widetilde{V}_0\right| \le \epsilon^{\rm QAE}_0 \label{eq:assQAEErrFin}
	\end{equation}
	is satisfied for some $\epsilon^{\rm QAE}_0\in\mathbb{R}_+$.
	Then,
	\begin{equation}
		|V_0-\widetilde{V}_0|\le \sum_{k=1}^{K-1} \widetilde{\Lambda}_{1,k-1} \epsilon^{\rm int}_{k} + \sum_{k=1}^{K-1} \widetilde{\Lambda}_{1,k-1} \epsilon^{\rm OB}_{k}+ \sum_{k=0}^{K-1} \widetilde{\Lambda}_{1,k} \epsilon^{\rm QAE}_{k} \label{eq:errOptPrice}
	\end{equation}
	holds, where, for $k\in[K-1]$ and $k^\prime\in[K-1]_0$,
	\begin{equation}
		\epsilon^{\rm int}_k := \epsilon_{\rm int}(\rho_{k},d,m_{k},B_{k})
	\end{equation}
	and
	\begin{eqnarray}
		\widetilde{\Lambda}_{k,k^\prime}&:=&
		\begin{cases}
			\prod_{i=k}^{k^\prime}\Lambda_i & ; \ {\rm if} \ k\le k^\prime \\
			1 & ; \ {\rm otherwise}
		\end{cases} \label{eq:LambdaTil} \\
		\Lambda_k &:=& \left(\frac{2}{\pi}\log(m_k+1)+1\right)^d. \label{eq:Lambda}
	\end{eqnarray}
	\label{th:err}
\end{theorem}
\noindent The proof is presented in Appendix \ref{sec:PrThErr}.

\subsection{Complexity}

Based on Theorem \ref{th:err}, we can evaluate the complexity, that is, the number of calls to $O_{{\rm step},k}$ sufficient to achieve the desired level of the error on the present option price.

\begin{corollary}
	Let $\epsilon$ be a real number in $(0,0.1)$.
	Under Assumptions \ref{ass:EvolOra} to \ref{ass:QkAnal}, consider Algorithm \ref{alg:new} with the following parameters:
	\begin{enumerate}
		\renewcommand{\labelenumi}{(\roman{enumi})}
		\item 
		$m_k,k\in[K-1]$ satisfying $m_k\ge m^{\rm th}_k$ with
		\begin{eqnarray}
			m^{\rm th}_1 &=& \left\lceil \frac{1}{\log\rho_1}\log\left(\frac{2^{d/2+2}\sqrt{d}(K-1)(1-\rho_1^{-2})^{-d/2}B_1}{\epsilon \widetilde{V}^{\rm max}_1}\right)\right\rceil \nonumber \\
			m^{\rm th}_k &=&  \left\lceil \frac{1}{\log\rho_k}\log\left(\frac{2^{d/2+2}\sqrt{d}(K-1)(1-\rho_k^{-2})^{-d/2}\widetilde{\Lambda}^{\rm th}_{1,k-1}B_k}{\epsilon \widetilde{V}^{\rm max}_1}\right)\right\rceil \nonumber \\
			&& \qquad\qquad\qquad\qquad\qquad\ {\rm for} \ k=2,...,K-1, \label{eq:mSuf}
		\end{eqnarray}
		where $\widetilde{\Lambda}^{\rm th}_{1,k-1}$ is determined as $\widetilde{\Lambda}_{1,k-1}$ in (\ref{eq:LambdaTil}) with $m_1=m^{\rm th}_1,...,m_{k-1}=m^{\rm th}_{k-1}$.
		
		\item
		$N^{\rm QAE}_k,k\in[K-1]_0$ set as
		\begin{equation}
			N^{\rm QAE}_k = \left\lceil \frac{7}{\bar{\epsilon}_k}\right\rceil. \label{eq:NQAESuf}
		\end{equation}
		Here, $\bar{\epsilon}_0,...,\bar{\epsilon}_{K-1}$ are given by
		\begin{eqnarray}
			\bar{\epsilon}_0 &=& \frac{1}{1+\sum_{k^\prime=1}^{K-1}\sqrt{(m_{k^\prime}+1)^d\widetilde{\Lambda}_{1,k^\prime}}}\cdot \frac{\epsilon}{4} \nonumber\\
			\bar{\epsilon}_k &=& \frac{\sqrt{(m_{k}+1)^d\big/\widetilde{\Lambda}_{1,k}}}{1+\sum_{k^\prime=1}^{K-1}\sqrt{(m_{k^\prime}+1)^d\widetilde{\Lambda}_{1,k^\prime}}} \cdot \frac{\widetilde{V}^{\rm max}_1 \epsilon}{4\widetilde{V}^{\rm max}_k}, \nonumber \\
			&&\qquad\qquad\qquad\qquad\qquad {\rm for} \ k=1,...,K-1,
		\end{eqnarray}
		where $m_0,...,m_{K-1}$ are set as (i) and $\widetilde{\Lambda}_{1,1},...,\widetilde{\Lambda}_{1,K-1}$ are given as (\ref{eq:LambdaTil}) with such $m_0,...,m_{K-1}$.
		
		\item
		$N^{\rm rep}_k$ set to
		\begin{equation}
			N_{\rm rep} := 12\left\lceil \log\left(\frac{N_{\rm est}}{0.01}\right)\right\rceil + 1, \label{eq:NRepSuf}
		\end{equation}
		for every $k\in[K-1]_0$.
		Here, $N_{\rm est}:=1+\sum_{k^\prime=1}^{K-1}(m_{k^\prime}+1)^d$ with $\{m_k\}$ set as (i).
	\end{enumerate}
	Moreover, suppose that $\epsilon^{\rm OB}_{1},...,\epsilon^{\rm OB}_{K-1}$ are 0.
	Then, Algorithm \ref{alg:new} outputs $\widetilde{V}_0$ satisfying $|V_0-\widetilde{V}_0|\le \epsilon \widetilde{V}^{\rm max}_1$ with probability higher than $0.99$.
	\label{col:comp}
\end{corollary}
\noindent The proof is presented in Appendix \ref{sec:PrColComp}.

We here explain why the parameters are set as above.
As we see in the proof in Appendix \ref{sec:PrColComp}, $m_1,...,m_{K-1}$ satisfying (\ref{eq:mSuf}) make the first term in the RHS in (\ref{eq:errOptPrice}) smaller than $\epsilon\widetilde{V}^{\rm max}_1/2$.
Then, for such $\{m_k\}_{k=1,...,K-1}$, $\{N^{\rm QAE}_k\}_{k=0,...,K-1}$ are determined as (\ref{eq:NQAESuf}) so that 
\begin{equation}
\frac{N_{\rm tot}}{N_{\rm rep}}=N^{\rm QAE}_0+\sum_{k=1}^{K-1}(m_k+1)^dN^{\rm QAE}_k, \label{eq:compPerEst}
\end{equation}
that is, the total number $N_{\rm tot}$ of calls to $\{O_{{\rm step},k}\}_{k=0,...,K-1}$ divided by the QAE repetition number $N^{\rm rep}$, is minimized under the constraint that, if all the QAEs in Algorithm \ref{alg:new} succeed, the third term in the RHS in (\ref{eq:errOptPrice}) is smaller than $\epsilon\widetilde{V}^{\rm max}_1/2$.
Finally, $\{N^{\rm rep}_k\}_{k=0,...,K-1}$ are determined so that the probability that these QAEs all succeed becomes higher than $0.99=1-0.01$.
In total, Algorithm \ref{alg:new} with the setting in Corollary \ref{col:comp} gives an approximation of $V_0$ with an error at most $\epsilon\widetilde{V}^{\rm max}_1$ with probability higher than $0.99$.

Note that, in reality, it is difficult to set $m_k$ to $m^{\rm th}_k$, since $\rho_k$ and $B_k$ are usually unknown.
In practice, we might set them to some conservatively large values, based on, for example, the calculation results of some benchmark pricing problems for various $\{m_k\}_{k=1,...,K-1}$.
Besides, note that, in the above setting, the half of the error tolerance $\epsilon$ is assigned to the interpolation error and another half is assigned to the QAE error.
Although we can of course change this assignment ratio, it affects the complexity only logarithmically, since the sufficient levels of $\{m_k\}_{k=1,...,K-1}$ are logarithmically affected by such a change and so are $\{N^{\rm rep}_k\}$ compensating the change of $\{m_k\}$.

Let us consider the dependency of the total complexity on the error tolerance $\epsilon$.
We see that
\begin{eqnarray}
	m_k&=&O\left(\log\left(\epsilon^{-1}){\rm polyloglog}(\epsilon^{-1}\right)\right), \\
	N^{\rm QAE}_k&=&O\left(\epsilon^{-1}\times{\rm polyloglog}(\epsilon^{-1})\right)
\end{eqnarray}
for every $k\in[K-1]$, and that
\begin{equation}
	N^{\rm QAE}_0=
	O\left(\epsilon^{-1}\log^{d/2}(\epsilon^{-1}){\rm polyloglog}(\epsilon^{-1})\right),
\end{equation}
where ${\rm polyloglog}(\cdot)$ means ${\rm polylog}\left(\log(\cdot)\right)$.
Combining these with (\ref{eq:compPerEst}), we obtain
\begin{equation}
	N_{\rm tot} = O\left(\epsilon^{-1}\log^{d}(\epsilon^{-1}){\rm polyloglog}(\epsilon^{-1})\right), \label{eq:Ntot}
\end{equation} 
which eventually beats LSM's complexity $\widetilde{O}(\epsilon^{-2})$ for small $\epsilon$.


\subsection{Comparison with existing Chebyshev interpolation-based methods \label{sec:ClCheb}}

In fact, the idea that we approximate the continuation value by Chebyshev interpolation is not novel.
There are some classical methods for Bermudan option pricing based on Chebyshev interpolation~\cite{Sullivan,Lim,Mahlstedt,Gas,Glau1,Glau2,Glau3}.
However, in addition to whether we use QAE or other classical methods for calculating the nodal continuation values, there are the following differences between the above proposed method and the existing methods.

First, we note that we do not have to use Monte Carlo for calculating the continuation value, and \cite{Sullivan,Lim} actually used other methods.
These papers considered the situation where the transition probability of the underlying asset prices can be easily calculated, and computed the continuation value by the numerical integration of the product of the transition probability and the option value at the next exercise date.
Note that this way is possible only for some simple models for underlying asset evolution such as the Black-Scholes model.
On the other hand, in more complicated settings where, for example, we price a multi-asset option under the stochastic local volatility model, Monte Carlo can be the sole solution, and such a time-consuming situation is a meaningful target for quantum speed-up.
However, combining Chebyshev interpolation with various methods might be an interesting possibility also in the quantum setup, and worth to be investigated as a future work.

Let us also mention the differences from \cite{Glau1}.
The major difference is that, in the method in \cite{Glau1}, the continuation value is not the target of either Monte Carlo integration or Chebyshev interpolation.
Instead, the method calculates the conditional expectations of Chebyshev polynomials by Monte Carlo or other methods, and find the Chebyshev interpolation of not the continuation value but the option price at each exercise date.
This approach saves the computational time when we price many options under a same model, since we can reuse the conditional expectations for interpolations in pricing different options. 
Considering the quantum version of this approach might be interesting too.

\subsection{Exponential factor with respect to the number of exercise dates in the error bound \label{sec:ExpComp}}

Now, let us make a comment on the factor $\widetilde{\Lambda}_{1,K-1}$ in (\ref{eq:errOptPrice}), which is reflected into the polyloglog factors in (\ref{eq:Ntot}).
This exponentially depends on the number of exercise dates $K$.
Therefore, it seems that the error on the option price exponentially grows as $K$ increase, and so does the complexity sufficient to achieve a given error tolerance.
Similar situations arose in the error analyses for LSM \cite{Glasserman2,Egloff,Zanger1,Zanger2,Zanger3,Zanger4} and classical Chebyshev interpolation-based methods \cite{Glau1}.

However, we should note that (\ref{eq:errOptPrice}) is an upper bound on the error, and that the actual error might not necessarily grows exponentially against $K$.
In fact, in the numerical experiment in \cite{Glau1}, where American options were approximately priced as Bermudan options with a small exercise date interval, the error was suppressed even if hundreds or thousands of exercise dates were set.

Let us now consider why the factor exponentially depending on $K$ appears in (\ref{eq:errOptPrice}).
In the derivation of (\ref{eq:errOptPrice}) described in Appendix \ref{sec:PrThErr}, we make Assumption \ref{ass:QkAnal} on the analyticity and boundedness of the continuation values $Q_k$, and apply Theorem \ref{th:ChebErrFunc} to Chebyshev interpolation of $Q_k$ in Algorithm \ref{alg:new}.
Since we use not $Q_k\left(\vec{S}^{\mathcal{D}_k,m_k}_{\vec{j}}\right)$, the values of $Q_k$ at the Chebyshev nodes, but $\widehat{Q}^{\rm QAE}_{k,\vec{j}}$, estimates on $Q_k\left(\vec{S}^{\mathcal{D}_k,m_k}_{\vec{j}}\right)$ with some errors, in interpolation, the term like the second term in the RHS in (\ref{th:ChebErrFunc}) arises in the upper bound of the difference between $Q_k$ and the interpolant $\widetilde{Q}_k$, and is amplified at every later interpolation. 

On the other hand, we can consider that the actual target of Chebyshev interpolation is not $Q_k$ but $\widehat{Q}_k$ in (\ref{eq:Qhat}).
That is, we can regard $\widehat{Q}^{\rm QAE}_{k,\vec{j}}$ as not an estimate on $Q_k\left(\vec{S}^{\mathcal{D}_k,m_k}_{\vec{j}}\right)$ but that on $\widehat{Q}_k\left(\vec{S}^{\mathcal{D}_k,m_k}_{\vec{j}}\right)$.
Then, we can make an assumption not on analyticity and boundedness of $Q_k$ but those of $\widehat{Q}_k$.
This leads to a different error bound than Theorem \ref{th:err}.
Actually, if we make Assumption \ref{ass:QhatkAnal} instead of Assumption \ref{ass:QkAnal}, we have an error bound with no exponential factor as shown in Theorem \ref{th:err2}.

\begin{assumption}
	For every $k\in[K-1]$, $\widehat{Q}_k(\vec{S})$ has an analytic extension to $\mathcal{B}_{\mathcal{D}_k,\widehat{\rho}_k}$, where $\mathcal{D}_k$ is given in Assumption \ref{ass:Extrap} and $\widehat{\rho}_k$ is some real number greater than 1, and
	\begin{equation}
		\sup_{\vec{s}\in \mathcal{B}_{\mathcal{D}_k,\widehat{\rho}_k}} \left|\widehat{Q}_k(\vec{S})\right| \le \widehat{B}_k
	\end{equation}
	holds, where $\widehat{B}_k$ is some positive real number.
	\label{ass:QhatkAnal}
\end{assumption}

\begin{theorem}
	Under Assumptions \ref{ass:EvolOra}, \ref{ass:Extrap}, and \ref{ass:QhatkAnal}, consider Algorithm \ref{alg:new}.
	Suppose that, for every $k\in[K-1]$ and $\vec{j}\in \mathcal{J}_k$, (\ref{eq:assQAEErr}) is satisfied for some $\epsilon^{\rm QAE}_k\in\mathbb{R}_+$, and that (\ref{eq:assQAEErrFin}) is satisfied for some $\epsilon^{\rm QAE}_0\in\mathbb{R}_+$.
	Then,
	\begin{equation}
		|V_0-\widetilde{V}_0|\le \sum_{k=1}^{K-1}\epsilon^{\rm OB}_{k}+\sum_{k=1}^{K-1}\widetilde{\epsilon}^{\rm int}_k + \sum_{k=0}^{K-1}\Lambda_k\epsilon^{\rm QAE}_k\label{eq:errOptPrice2}
	\end{equation}
	holds, where, for every $k\in[K-1]$,
	\begin{equation}
		\widetilde{\epsilon}^{\rm int}_k := \epsilon_{\rm int}(\widetilde{\rho}_{k},d,m_{k},\widetilde{B}_{k}),
	\end{equation}
	and $\Lambda_k$ is defined as (\ref{eq:Lambda}).
	\label{th:err2}
\end{theorem}
\noindent The proof is presented in Appendix \ref{sec:PrThErr2}.
Note that a similar point has been made for LSM in \cite{Zanger2} (Theorem 3.1).

Of course, $\widehat{Q}_k$ is defined as (\ref{eq:Qhat}) with $\widetilde{V}_{k+1}$, which is the intermediate output in Algorithm \ref{alg:new}, and making assumptions on such a thing does not lead to self-contained discussion.
It is more desirable to derive the error bound under assumptions on $Q_k$ and/or other quantities determined independently from pricing algorithms.
We leave considering whether we can obtain an error bound similar to Theorem \ref{th:err2} under such assumptions or not as a future work.

\subsection{Quantization of LSM \label{sec:QLSM}}

Lastly, we make a comment on whether we can consider the quantum algorithm for LSM.
Since there are some quantum algorithms for linear regression\cite{Wiebe,Schuld,Wang2,Yu,Chakraborty,Kerenidis3,Kaneko2}, we naturally wonder that we can apply these to LSM and then obtain speed-up.
However, this is not so straightforward, since most of these algorithms output the regression result as a quantum state, in which the regression coefficients are amplitude-encoded.
Fortunately, some algorithms \cite{Wang2,Kaneko2} output the regression coefficient as classical data.
In particular, the algorithm in \cite{Kaneko2} has the complexity of $\widetilde{O}(D^{7/2}\kappa^4/\epsilon)$ with the tolerance $\epsilon$, the explanatory variable number $D$, and the condition number $\kappa$ of the design matrix, from which we expect the quadratic speed-up of LSM with respect to $\epsilon$.
Nevertheless, applying this algorithm to LSM is not immediate either, because of some points to be considered.
For example, the complexity has strong dependence on $D$ and $\kappa$, which might make the algorithm disadvantageous.
Therefore, it will be crucial to evaluate these, especially $\kappa$, in addition to finding the basis function set which makes $\kappa$ as small as possible.
We will consider this direction in the future work.

\section{\label{sec:Sum}Summary}

In this paper, we considered application of quantum algorithms to Bermudan option pricing.
Since there are QAE-based algorithms for Monte Carlo integration, which provide quadratic speed-up compared with the classical counterparts, and applications of them to some option pricing problems have been investigated, it is natural to consider to apply them to Bermudan option pricing.
One crucial issue in this problem is how to approximate the continuation value $Q_k$, which determines the optimal exercise date.
In order to cope with this, we considered combination of QAE and Chebyshev interpolation.
That is, the proposed method estimates the values of $Q_k$ on the interpolation nodes by QAE, and, using such estimates, find a Chebyshev interpolation as an approximation of $Q_k$.
We presented the calculation procedure in detail, along with the error bound and the complexity, which corresponds to the number of calls to the oracle for underlying asset price evolution, sufficient to achieve the desired error tolerance $\epsilon$.
As expected, this method has the complexity depending on $\epsilon$ as $\widetilde{O}(\epsilon^{-1})$, which means the quadratic speed-up compared with LSM, the typical classical algorithm for Bermudan option pricing.

As a future work, it is interesting to consider the quantum version of LSM, as mentioned in Section \ref{sec:QLSM}.
Besides, it is also meaningful to extend the proposed method to other types of dynamic programming, which is ubiquitous in many fields of science and industry.

\section*{Acknowledgment}

This work was supported by MEXT Quantum Leap Flagship Program (MEXT Q-LEAP) Grant Number JPMXS0120319794.

\appendix

\section{\label{sec:PrThErr} Proof of Theorem \ref{th:err}}

\begin{proof}
	First, we note that, for every $k\in[K-1]$,
	\begin{equation}
		\epsilon_k \le \tilde{\epsilon}_k \label{eq:epskIter}
	\end{equation}
	holds, where
	\begin{equation}
		\epsilon_k:=
		\max_{\vec{S}\in\mathcal{D}_k} |V_k(\vec{S})-\widetilde{V}_k(\vec{S})|, \label{eq:epsk}
	\end{equation}
	and
	\begin{equation}
		\tilde{\epsilon}_k:=
		\begin{cases}
			\epsilon^{\rm int}_{K-1} + \Lambda_{k}\epsilon^{\rm QAE}_{K-1} & ; \ {\rm for} \ k=K-1 \\
			\epsilon^{\rm int}_k + \Lambda_{k}\left(\epsilon_{k+1} + \epsilon^{\rm OB}_{k+1} + \epsilon^{\rm QAE}_k\right) & ;  \ {\rm for} \ k = 1,...,K-2
		\end{cases}.
	\end{equation}
	The proof of this is as follows.
	We see that, for any $k\in[K-1]$,
	\begin{equation}
		\left|V_{k}(\vec{S}_{k}) - \mathbf{G}_{k}[\widetilde{V}_{k}](\vec{S}_{k})\right| 
		= \left|V_{k}(\vec{S}_{k}) - \widetilde{V}_{k}(\vec{S}_{k})\right|
		\le \epsilon_k 
	\end{equation}
	holds if $\vec{S}_{k}\in\mathcal{D}_k$, and, under Assumption \ref{ass:Extrap}, either
	\begin{equation}
		\left|V_{k}(\vec{S}_{k}) - \mathbf{G}_{k}[\widetilde{V}_{k}](\vec{S}_{k})\right| 
		= \left|V_{k}(\vec{S}_{k}) - V^{\rm OB}_{k}(\vec{S}_{k})\right|
		\le \epsilon^{\rm OB}_k
	\end{equation}
	or
	\begin{eqnarray}
		&&\left|V_{k}(\vec{S}_{k}) - \mathbf{G}_{k}[\widetilde{V}_{k}](\vec{S}_{k})\right| \nonumber \\
		&=& \left|V_{k}(\vec{S}_{k}) - \mathbf{F}_{k}[\widetilde{V}_{k}](\vec{S}_{k})\right| \nonumber\\
		&\le& \left|V_{k}(\vec{S}_{k}) - \mathbf{F}_{k}[V_{k}](\vec{S}_{k})\right| +\left|\mathbf{F}_{k}[V_{k}](\vec{S}_{k}) - \mathbf{F}_{k}[\widetilde{V}_{k}](\vec{S}_{k})\right| \nonumber \\
		&=& \left|V_{k}(\vec{S}_{k}) - \mathbf{F}_{k}[V_{k}](\vec{S}_{k})\right| +\left|V_{k}(b_k(\vec{S}_{k})) - \widetilde{V}_{k}(b_k(\vec{S}_{k}))\right| \nonumber \\
		&\le&\epsilon^{\rm OB}_k + \epsilon_k 
	\end{eqnarray}
	holds if $\vec{S}_{k}\in\mathcal{S}\setminus\mathcal{D}_k$.
	Combining these, we obtain
	\begin{equation}
		\left|V_{k}(\vec{S}_{k}) - \mathbf{G}_{k}[\widetilde{V}_{k}](\vec{S}_{k})\right| \le \epsilon_{k} + \epsilon^{\rm OB}_k \label{eq:diffVandGV}
	\end{equation}
	for any $\vec{S}_{k}\in\mathcal{S}$.
	This leads to
	\begin{eqnarray}
		&&|Q_k(\vec{S})-\widehat{Q}_k(\vec{S})| \nonumber \\
		&=& \left|\mathbb{E}\left[V_{k+1}(\vec{S}_{k+1}) - \mathbf{G}_{k+1}[\widetilde{V}_{k+1}](\vec{S}_{k+1}) \ \middle| \vec{S}_{k}=\vec{S}\right]\right| \nonumber \\
		&\le& \mathbb{E}\left[\left|V_{k+1}(\vec{S}_{k+1}) - \mathbf{G}_{k+1}[\widetilde{V}_{k+1}](\vec{S}_{k+1})\right| \ \middle| \vec{S}_{k}=\vec{S}\right] \nonumber \\
		&\le&\epsilon_{k+1} + \epsilon^{\rm OB}_{k+1} \label{eq:temp2}
	\end{eqnarray}
	for any $k\in[K-2]$ and $\vec{S}\in\mathcal{D}_k$.
	Thus, with (\ref{eq:assQAEErr}), we obtain
	\begin{eqnarray}
		&&\left|Q_k\left(\vec{S}^{\mathcal{D}_k,m_k}_{\vec{j}}\right)-\widehat{Q}^{\rm QAE}_{k,\vec{j}}\right| \nonumber \\
		&\le& \left|Q_k\left(\vec{S}^{\mathcal{D}_k,m_k}_{\vec{j}}\right)-\widehat{Q}_k\left(\vec{S}^{\mathcal{D}_k,m_k}_{\vec{j}}\right)\right| + \left|\widehat{Q}_k\left(\vec{S}^{\mathcal{D}_k,m_k}_{\vec{j}}\right)-\widehat{Q}^{\rm QAE}_{k,\vec{j}}\right| \nonumber \\
		&\le& \epsilon_{k+1} + \epsilon^{\rm OB}_{k+1} + \epsilon^{\rm QAE}_k
	\end{eqnarray}
	for every $k\in[K-2]$ and $\vec{j}\in\mathcal{J}_k$.
	On the other hand, for $k=K-1$, noting that $Q_{K-1}(\vec{S})=\widehat{Q}_{K-1}(\vec{S})$ for any $\vec{S}\in\mathcal{S}$ by definition, we see that
	\begin{equation}
		\left|Q_{K-1}\left(\vec{S}^{\mathcal{D}_{K-1},m_{K-1}}_{\vec{j}}\right)-\widehat{Q}^{\rm QAE}_{K-1,\vec{j}}\right| \le \epsilon^{\rm QAE}_{K-1}
	\end{equation}
	for any $\vec{j}\in\mathcal{J}_{K-1}$.
	Then, under Assumption \ref{ass:QkAnal}, invoking Theorem \ref{th:ChebErrFunc}, we obtain
	\begin{equation}
		|Q_k(\vec{S})-\widetilde{Q}_k(\vec{S})| \le \tilde{\epsilon}_k 
	\end{equation}
	for any $k\in[K-1]$ and $\vec{S}\in\mathcal{D}_k$, which immediately leads to (\ref{eq:epskIter}) as
	\begin{eqnarray}
		&&\left|V_k(\vec{S})-\widetilde{V}_k(\vec{S})\right| \nonumber \\
		&=& \left|\max\{f^{\rm pay}_k(\vec{S}),Q_k(\vec{S})\} - \max\{f^{\rm pay}_k(\vec{S}),\widetilde{Q}_k(\vec{S})\}\right| \nonumber \\
		&\le& \left|Q_k(\vec{S})-\widetilde{Q}_k(\vec{S})\right| \nonumber \\
		&\le& \tilde{\epsilon}_k. \label{eq:VkDiffUB}
	\end{eqnarray}
	Here, we used $|\max\{a,b\}-\max\{a,c\}|\le|b-c|$, which holds for any $a,b,c\in\mathbb{R}$.\\
	
	Next, let us note that, for $k\in[K-2]$,
	\begin{equation}
		\epsilon_k \le \sum_{k^\prime=k}^{K-1} \widetilde{\Lambda}_{k,k^\prime-1} \epsilon^{\rm int}_{k^\prime} + \sum_{k^\prime=k+1}^{K-1} \widetilde{\Lambda}_{k,k^\prime-1} \epsilon^{\rm OB}_{k^\prime}+ \sum_{k^\prime=k}^{K-1} \widetilde{\Lambda}_{k,k^\prime} \epsilon^{\rm QAE}_{k^\prime}, \label{eq:epskUB}
	\end{equation}
	holds.
	We prove this by induction.
	For $k=K-2$, (\ref{eq:epskIter}) implies that
	\begin{eqnarray}
		&&\epsilon_{K-2} \nonumber \\
		&\le& \epsilon^{\rm int}_{K-2} + \Lambda_{K-2}\left(\epsilon_{K-1}+\epsilon^{\rm OB}_{K-1}+\epsilon^{\rm QAE}_{K-2}\right) \nonumber \\
		&\le& \epsilon^{\rm int}_{K-2} + \Lambda_{K-2}\left(\epsilon^{\rm int}_{K-1} + \Lambda_{K-1}\epsilon^{\rm QAE}_{K-1}+\epsilon^{\rm OB}_{K-1}+\epsilon^{\rm QAE}_{K-2}\right) \nonumber \\
		&=& \sum_{k^\prime=K-2}^{K-1} \widetilde{\Lambda}_{K-2,k^\prime-1} \epsilon^{\rm int}_{k^\prime} + \sum_{k^\prime=K-1}^{K-1} \widetilde{\Lambda}_{K-2,k^\prime-1} \epsilon^{\rm OB}_{k^\prime}+ \sum_{k^\prime=K-2}^{K-1} \widetilde{\Lambda}_{K-2,k^\prime} \epsilon^{\rm QAE}_{k^\prime} \nonumber \\
		&&
	\end{eqnarray}
	Similarly, if (\ref{eq:epskUB}) hold for $k\in\{2,...,K-2\}$, (\ref{eq:epskIter}) implies that
	\begin{eqnarray}
		&&\epsilon_{k-1}\nonumber \\
		&\le& \epsilon^{\rm int}_{k-1} + \Lambda_{k-1}\left(\epsilon_{k}+\epsilon^{\rm OB}_{k}+\epsilon^{\rm QAE}_{k-1}\right) \nonumber \\
		&\le& \epsilon^{\rm int}_{k-1} + \Lambda_{k-1}\Bigg(\sum_{k^\prime=k}^{K-1} \widetilde{\Lambda}_{k,k^\prime-1} \epsilon^{\rm int}_{k^\prime} + \sum_{k^\prime=k+1}^{K-1} \widetilde{\Lambda}_{k,k^\prime-1} \epsilon^{\rm OB}_{k^\prime}+ \sum_{k^\prime=k}^{K-1} \widetilde{\Lambda}_{k,k^\prime} \epsilon^{\rm QAE}_{k^\prime}  \nonumber \\
		&& \qquad\qquad\qquad\qquad\qquad\qquad\qquad\qquad +\epsilon^{\rm OB}_{k}+\epsilon^{\rm QAE}_{k-1}\Bigg) \nonumber \\
		&=& \sum_{k^\prime=k-1}^{K-1} \widetilde{\Lambda}_{k-1,k^\prime-1} \epsilon^{\rm int}_{k^\prime} + \sum_{k^\prime=k}^{K-1} \widetilde{\Lambda}_{k-1,k^\prime-1} \epsilon^{\rm OB}_{k^\prime}+ \sum_{k^\prime=k-1}^{K-1} \widetilde{\Lambda}_{k-1,k^\prime} \epsilon^{\rm QAE}_{k^\prime}. \nonumber \\
		&&
	\end{eqnarray}
	Therefore, (\ref{eq:epskUB}) is proved for every $k\in[K-2]$.\\
	
	Finally, the claim is proved as follows.
	We see that
	\begin{eqnarray}
		&& |V_0-\widehat{V}_0| \nonumber \\
		&=& \left|\mathbb{E}[V_1(\vec{S}_1)]-\mathbb{E}[\mathbf{G}_1[\tilde{V}_1](\vec{S}_1)]\right|\nonumber \\
		&\le& \mathbb{E}\left[\left|V_1(\vec{S}_1)-\mathbf{G}_1[\tilde{V}_1](\vec{S}_1)\right|\right]\nonumber \\
		&\le& \epsilon_1+\epsilon^{\rm OB}_1 \nonumber \\
		&\le& \sum_{k=1}^{K-1} \widetilde{\Lambda}_{1,k-1} \epsilon^{\rm int}_{k} + \sum_{k=1}^{K-1} \widetilde{\Lambda}_{1,k-1} \epsilon^{\rm OB}_{k}+ \sum_{k=1}^{K-1} \widetilde{\Lambda}_{1,k} \epsilon^{\rm QAE}_{k}, \label{eq:V0DiffBound}
	\end{eqnarray}
	where we used (\ref{eq:diffVandGV}) and (\ref{eq:epskUB}) at the second and last inequalities, respectively.
	Combining this and (\ref{eq:assQAEErrFin}), we obtain
	\begin{eqnarray}
		&&|V_0-\widetilde{V}_0| \nonumber \\
		&\le&|V_0-\widehat{V}_0| + |\widehat{V}_0-\widetilde{V}_0| \nonumber \\
		&=& \sum_{k=1}^{K-1} \widetilde{\Lambda}_{1,k-1} \epsilon^{\rm int}_{k} + \sum_{k=1}^{K-1} \widetilde{\Lambda}_{1,k-1} \epsilon^{\rm OB}_{k}+ \sum_{k=0}^{K-1} \widetilde{\Lambda}_{1,k} \epsilon^{\rm QAE}_{k}.
	\end{eqnarray}
\end{proof}

\section{Proof of Corollary \ref{col:comp} \label{sec:PrColComp}}

\begin{proof}
	By simple algebra, we see that $m_1,...,m_{K-1}$ satisfying (\ref{eq:mSuf}) lead to
	\begin{equation}
		\widetilde{\Lambda}_{1,k-1}\epsilon^{\rm int}_k \le \frac{\epsilon \widetilde{V}^{\rm max}_1}{2(K-1)} \label{eq:intErr}
	\end{equation}
	for every $k\in[K-1]$.
	
	On the other hand, Theorem \ref{th:QAE} implies that, for every $k\in[K-1]$ and $\vec{j}\in\mathcal{J}_k$, using $N^{\rm QAE}_k$ set as (\ref{eq:NQAESuf}), Step 4 in Algorithm \ref{alg:new} gives us $\widetilde{P}_{k,\vec{j}}$, an estimate on $P_{k,\vec{j}}$ in (\ref{eq:Pkj}), satisfying
	\begin{equation}
		\left|P_{k,\vec{j}}-\widetilde{P}_{k,\vec{j}}\right| \le \bar{\epsilon}_k, \label{eq:PkErrAct}
	\end{equation}
	and then $\widehat{Q}^{\rm QAE}_{k,\vec{j}}$ satisfying
	\begin{eqnarray}
		&&\left|\widehat{Q}_{k}\left(\vec{S}^{\mathcal{D}_{k},m_{k}}_{\vec{j}}\right)-\widehat{Q}^{\rm QAE}_{k,\vec{j}}\right| \nonumber \\
		&=& \left|(2P_{k,\vec{j}}-1)\widetilde{V}^{\rm max}_k-(2\widetilde{P}_{k,\vec{j}}-1)\widetilde{V}^{\rm max}_k\right| \nonumber \\
		&\le& 2\widetilde{V}^{\rm max}_k \bar{\epsilon}_k \nonumber \\
		&\le& \frac{\sqrt{(m_{k}+1)^d\big/\widetilde{\Lambda}_{1,k}}}{1+\sum_{k^\prime=1}^{K-1}\sqrt{(m_{k^\prime}+1)^d\widetilde{\Lambda}_{1,k^\prime}}} \cdot \frac{\widetilde{V}^{\rm max}_1 \epsilon}{2}=:\widetilde{\epsilon}^{\rm QAE}_k, \label{eq:QkErrAct}
	\end{eqnarray}
	with some probability.
	Similarly, it is implied that, with $N^{\rm QAE}_0$ set as (\ref{eq:NQAESuf}), Step 9 
	gives us $\widetilde{P}_{0}$, an estimation on $P_{0}$, satisfying
	\begin{equation}
		\left|P_0-\widetilde{P}_0\right| \le \bar{\epsilon}_0, \label{eq:P0ErrAct}
	\end{equation}
	and then $\widetilde{V}_{0}$ satisfying
	\begin{eqnarray}
		&&\left|\widehat{V}_{0}-\widetilde{V}_{0}\right| \nonumber \\
		&=& \left|(2P_0-1)\widetilde{V}^{\rm max}_0-(2\widetilde{P}_0-1)\widetilde{V}^{\rm max}_0\right| \nonumber \\
		&\le& 2\widetilde{V}^{\rm max}_0 \bar{\epsilon}_0 \nonumber \\
		&\le& \frac{1}{1+\sum_{k^\prime=1}^{K-1}\sqrt{(m_{k^\prime}+1)^d\widetilde{\Lambda}_{1,k^\prime}}} \cdot \frac{\widetilde{V}^{\rm max}_1 \epsilon}{2}=:\widetilde{\epsilon}^{\rm QAE}_0, \label{eq:V0ErrAct}
	\end{eqnarray}
	with some probability.
	Therefore, when all of these succeed,
	\begin{eqnarray}
		&& |V_0-\widetilde{V}_0| \nonumber \\
		&\le& \sum_{k=1}^{K-1} \widetilde{\Lambda}_{1,k-1} \epsilon^{\rm int}_{k} + \sum_{k=1}^{K-1} \widetilde{\Lambda}_{1,k-1} \epsilon^{\rm OB}_{k}+ \sum_{k=0}^{K-1} \widetilde{\Lambda}_{1,k} \widetilde{\epsilon}^{\rm QAE}_{k} \nonumber \\
		&\le& \frac{\epsilon \widetilde{V}^{\rm max}_1}{2} + 0 +\frac{\epsilon \widetilde{V}^{\rm max}_1}{2} \nonumber \\
		&=& \epsilon \widetilde{V}^{\rm max}_1
	\end{eqnarray}
	holds, according to Theorem \ref{th:err}. 
	Here, we used (\ref{eq:intErr}), (\ref{eq:QkErrAct}), (\ref{eq:V0ErrAct}), and the assumption that $\epsilon^{\rm OB}_1,...,\epsilon^{\rm OB}_{K-1}$ are 0, along with simple algebra.
	
	The remaining task is proving that the probability $P_{\rm all}$ that all of the estimations in Steps 4 and 9 succeed is larger than 0.99 under the setting of $N^{\rm rep}_k$ as (\ref{eq:NRepSuf}).
	Note that, according to Theorem \ref{th:QAE}, with the setting as (\ref{eq:NRepSuf}), the probability that Step 4 for a set of $k\in[K-1]$ and $\vec{j}\in\mathcal{J}_k$ outputs $\widetilde{P}_{k,\vec{j}}$ satisfying (\ref{eq:PkErrAct}) is higher than
	\begin{equation}
		1-\frac{0.01}{N_{\rm est}}. \label{eq:sucProbOne}
	\end{equation}
	Similarly, the probability that Step 9 outputs $\widetilde{P}_0$ satisfying (\ref{eq:P0ErrAct}) is also higher than (\ref{eq:sucProbOne}).
	Besides, the total number of these estimations is $N_{\rm est}$.
	Combining these, we obtain a lower bound of $P_{\rm all}$ as
	\begin{equation}
		P_{\rm all}\ge \left(1-\frac{0.01}{N_{\rm est}}\right)^{N_{\rm est}}\ge 1-0.01 = 0.99,
	\end{equation}
	which completes the proof.
\end{proof}

\section{\label{sec:PrThErr2} Proof of Theorem \ref{th:err2}}

\begin{proof}
	Because of (\ref{eq:assQAEErr}), Theorem \ref{th:ChebErrFunc} implies that
	\begin{equation}
		\left|\widehat{Q}_k(\vec{S})-\widetilde{Q}_k(\vec{S})\right| \le \widetilde{\epsilon}^{\rm int}_k + \Lambda_k\epsilon^{\rm QAE}_k \label{eq:tocyuu1}
	\end{equation}
	for every $k\in[K-1]$ and $\vec{S}\in\mathcal{D}_k$.
	Besides, for every $k\in[K-2]$ and $\vec{S}\in\mathcal{D}_k$, $\left|Q_k(\vec{S})-\widehat{Q}_k(\vec{S})\right|\le\epsilon_{k+1}+\epsilon^{\rm OB}_{k+1}$ 
	holds as (\ref{eq:temp2}), where $\epsilon_{k}$ is defined as (\ref{eq:epsk}) for every $k\in[K-1]$, whereas $Q_{K-1}(\vec{S})=\widehat{Q}_{K-1}(\vec{S})$ by definition of $\mathbf{G}_K[\cdot]$.
	Combining these, we see that, for every $k\in[K-1]$ and $\vec{S}\in\mathcal{D}_k$,
	\begin{eqnarray}
		\left|Q_k(\vec{S})-\widetilde{Q}_k(\vec{S})\right| &\le& \left|Q_k(\vec{S})-\widehat{Q}_k(\vec{S})\right| + \left|\widehat{Q}_k(\vec{S})-\widetilde{Q}_k(\vec{S})\right| \nonumber \\
		&\le& \epsilon_{k+1}+\epsilon^{\rm OB}_{k+1}+\widetilde{\epsilon}^{\rm int}_k + \Lambda_k\epsilon^{\rm QAE}_k
	\end{eqnarray}
	holds with $\epsilon_{K}=0$ and $\epsilon^{\rm OB}_K=0$, which leads to
	\begin{equation}
		\left|V_k(\vec{S})-\widetilde{V}_k(\vec{S})\right| \le \epsilon_{k+1}+\epsilon^{\rm OB}_{k+1}+\widetilde{\epsilon}^{\rm int}_k + \Lambda_k\epsilon^{\rm QAE}_k
	\end{equation}
	similarly to (\ref{eq:VkDiffUB}).
	Therefore, for every $k\in[K-1]$,
	\begin{equation}
		\epsilon_{k} =\max_{\vec{S}\in\mathcal{D}_k} |V_k(\vec{S})-\widetilde{V}_k(\vec{S})| \le \epsilon_{k+1}+\epsilon^{\rm OB}_{k+1}+\widetilde{\epsilon}^{\rm int}_k + \Lambda_k\epsilon^{\rm QAE}_k.
	\end{equation}
	This implies
	\begin{equation}
		\epsilon_{1} \le \sum_{k=2}^{K-1}\epsilon^{\rm OB}_{k}+\sum_{k=1}^{K-1}\widetilde{\epsilon}^{\rm int}_k + \sum_{k=1}^{K-1}\Lambda_k\epsilon^{\rm QAE}_k. \label{eq:tocyuu3}
	\end{equation}
	Finally, combining (\ref{eq:tocyuu3}) with (\ref{eq:assQAEErrFin}) and $|V_0-\widehat{V}_0|\le \epsilon_1+\epsilon^{\rm OB}_1$, which we can see as (\ref{eq:V0DiffBound}), we obtain
	\begin{eqnarray}
		&&|V_0-\widetilde{V}_0| \nonumber \\
		&\le& |V_0-\widehat{V}_0| + |\widehat{V}_0-\widetilde{V}_0| \nonumber \\
		&\le& \epsilon_1+\epsilon^{\rm OB}_1 + \epsilon^{\rm QAE}_0 \nonumber \\
		&\le& \sum_{k=1}^{K-1}\epsilon^{\rm OB}_{k}+\sum_{k=1}^{K-1}\widetilde{\epsilon}^{\rm int}_k + \sum_{k=0}^{K-1}\Lambda_k\epsilon^{\rm QAE}_k.
	\end{eqnarray}
\end{proof}

\end{document}